# Speciation and introgression between *Mimulus nasutus* and *Mimulus guttatus*

Yaniv Brandvain [1,§,*], Amanda M. Kenney [2,§], Lex Flagel[3], Graham Coop[1,⌘], and Andrea L. Sweigart[2,⌘]

1 Department of Evolution and Ecology & Center for Population Biology, University of California - Davis, Davis, CA, USA

2 Department of Genetics, University of Georgia, Athens, GA, USA

3 Monsanto Company, Chesterfield, MO, USA

§ These authors contributed equally to this work
⌘ These authors jointly supervised this work
* Author for correspondence – Email: ybrandvain@gmail.com

**ABSTRACT**

*Mimulus guttatus* and *M. nasutus* are an evolutionary and ecological model sister species pair differentiated by ecology, mating system, and partial reproductive isolation. Despite extensive research on this system, the history of divergence and differentiation in this sister pair is unclear. We present and analyze a novel population genomic data set which shows that *M. nasutus* ``budded'' off of a central Californian *M. guttatus* population within the last 200 to 500 thousand years. In this time, the *M. nasutus* genome has accrued numerous genomic signatures of the transition to predominant selfing. Despite clear biological differentiation, we document ongoing, bidirectional introgression. We observe a negative relationship between the recombination rate and divergence between *M. nasutus* and sympatric *M. guttatus* samples, suggesting that selection acts against *M. nasutus* ancestry in *M. guttatus*.

While speciation is often depicted as a simple event in which a single species splits into two, there is increasing evidence that this process is often more complex. In particular, speciation reflects a tension among divergence, the assortment of ancestral variation, and introgression that plays out across the geography and ecology of the incipient species. Our population genetic view of this process has been fundamentally limited by examining few loci, where stochasticity in ancestral processes can prevent strong inferences about isolation and gene flow. By contrast, whole genome resequencing (even of only a few individuals) makes use of both many genealogical histories and contiguous genomic blocks of ancestry to provide well-resolved views of population history, divergence and introgression (1-4). Here, we present a population genomic investigation of the speciation history of two closely related species of yellow monkeyflowers, the primarily outcrossing *Mimulus guttatus*, and the self-pollinating *M. nasutus* – an evolutionary model system for which the genetic and ecological basis of reproductive isolation is reasonably well characterized (5).

In flowering plants, speciation often involves a shift in pollinator (*e.g.*, (6-8)) or mating system (*e.g.*, (6, 9-11)), with concomitant divergence in key floral traits causing reproductive isolation between lineages. The evolutionary transition from outcrossing to self-fertilization, as occurred in *M. nasutus*, is of particular interest because the expected reduction in both the effective population size and effective recombination rate (12, 13) can dramatically alter population genetic processes, and patterns of genomic variation (14, 15). Recent evidence for elevated levels of putatively deleterious alleles in selfing taxa (16-18) is consistent with the idea that inbreeding reduces the effectiveness of purifying selection (due to a lowered effective population size). However, we still have few examples of the effects of self-fertilization on patterns of diversity across the genome, particularly in the context of recently diverged and potentially hybridizing species. Genomic datasets from young selfing species can uniquely inform the process of mating system divergence by allowing us to compare regions of the genome that share a common ancestor before or after the origin of self-fertilization and thus understand the assortment of ancestral variation (19).

The *M. guttatus-M. nasutus* species pair is an excellent model for investigating the causes and consequences of mating system evolution and species divergence. *M. guttatus* is primarily outcrossing (although the rate of outcrossing varies among populations, 20, 21, 22) with large, bee-pollinated flowers and occupies diverse ecological habitats throughout western North America. *M. nasutus* is highly selfing with reduced, mostly closed flowers and overlaps broadly with *M. guttatus* throughout their shared range. Though allopatric populations are more common, the two species often co-occur. In sympatry, species are partially reproductively isolated by differences in floral morphology, flowering phenology, and pollen-pistil interactions (23-25). Although early-generation hybrids occur in nature (23, 26), numerous intrinsic hybrid incompatibilities decrease hybrid fitness (27-29). Based on the most detailed population genetic analyses of *Mimulus* to date (two and six sequenced nuclear loci, respectively: 29, 30), *M. nasutus* exhibits reduced diversity compared to *M. guttatus,* and some *M. guttatus* sequences are nearly identical to *M. nasutus*, suggestive of historical introgression. However, this very limited view of the genome cannot resolve the timing and genomic consequences of divergence between *Mimulus* species, nor can it inform the extent or consequences of introgression between them.

We present the first population genomic analysis of *M. guttatus* and *M. nasutus*, spanning diverse ecotypes collected from throughout the species' ranges. We use these dense and contiguous population genomic data to estimate the population-split time, quantify rapid loss of ancestral variation accompanying the transition to selfing in *M. nasutus*, and identify ongoing, bidirectional introgression. Additionally, we observe a negative correlation between the recombination rate and interspecific divergence between *M. nasutus* and sympatric *M. guttatus*, a result best explained by selection against introgression of *M. nasutus* ancestry. Our approach provides a detailed view of differentiation and introgression in a tractable ecological, genetic, and evolutionary model system.

**Results**

We analyze a novel population genomic dataset of nineteen lab and/or naturally inbred *Mimulus* samples – thirteen *M. guttatus*, five *M. nasutus*, and one from the outgroup, *M. dentilobus*. Collections spanned the ecological and geographic ranges of each species (Figure 1A and Table S1). Many of our analyses focus on four *M. guttatus* and four *M. nasutus* collections sequenced to high depth (13.8X-24.7X) and with identical read lengths (100 bp, paired-end reads). All comparisons are presented in supplementary tables S2 and S3. Of our focal *M. guttatus* samples, CACG and DPRG are narrowly sympatric with *M. nasutus* populations from the northern and southern portion of both species' ranges, respectively. Our focal northern allopatric *M. guttatus* collection, AHQT, is well outside the geographic range of *M. nasutus*. By contrast, the southern allopatric collection (SLP) is geographically close to *M. nasutus* populations. Our focal *M. nasutus* samples also include sympatric and allopatric samples from the north and south (Table S1).

*Speciation history:*

Overall patterns of genomic differentiation show deep population structure in *M. guttatus*, with *M. nasutus* diverging from a central Californian *M. guttatus* population approximately 200 kya.

A rate-smoothed neighbor-joining (nj) tree clearly displays a deep phylogenetic split within *M. guttatus*, roughly corresponding to northern and southern parts of its range, and the placement of all *M. nasutus* samples as a node within the southern *M. guttatus* cluster (Figure 1B). Note, however, that geography is not a clean predictor of genetic structure within *M. guttatus* (*e.g.*, DUN is from a northern latitude yet clusters with our southern *M. guttatus* samples). The fact that *M. guttatus* is paraphyletic (*i.e.*, is split by *M. nasutus*), suggests that *M. nasutus* budded from within a structured ancestral *M. guttatus* population. A principle component analysis (PCA, Figure 1C) also reveals the genetic structure within *M. guttatus* as PC2 differentiates northern and southern *M. guttatus* groups identified in the nj tree. Consistent with the single origin of *M. nasutus*, PC1 separates *M. guttatus* from the strongly clustered *M. nasutus*, presumably as a consequence of a shared history of genetic drift among these *M. nasutus* samples,

We generate a quantitative description of the strong genetic structure within *M. guttatus*, focusing on our high-coverage (focal) samples. Pairwise sequence diversity at synonymous sites within northern ($\pi_{S\ AHQT\ x\ CACG}$ = 3.97% [3.89%-4.06%]) and southern ($\pi_{S\ DPR\ x\ SLP}$ = 4.45% [4.39%-4.52%]) samples is significantly lower than that within *M. guttatus* overall ($\pi_S$ = 4.91% [4.85%-4.96%]), or between north and south ($\pi_S$= 5.26% [5.20%-5.30%], Figure 1D). Diversity within the northern and southern clades is consistent with a very large effective population size ($N_e$) of approximately one and a half million chromosomes for both groups (assuming the per generation mutation rate, $\mu$ = 1.5 * $10^{-8}$ [following Koch *et al.* 2008], 2 $N_e$ = 1.5 x $10^6$, and assuming an annual life-history this is also the per year mutation rate). One simple way to estimate a population split time ($\tau$ generations), assuming no introgression, is to assume that the divergence between populations is the sum of pairwise diversity ($\pi$) within an ancestral population and the product of the per-generation mutation rate, $\mu$, and two times the split time (31). Using this relationship, and representing ancestral diversity by the southern *M. guttatus* samples, we set $\tau$= ($\pi_{S\ NorthGut\ x\ SouthGut}$ - $\pi_{S\ SouthGut}$) / 2$\mu$ and estimate a split between northern and southern *Mimulus* populations more than a quarter of a million years ago (265 ky [251 ky – 280 ky]).

Interspecific divergence between *M. guttatus* and *M. nasutus* ($d_S$ = 4.94% [4.88%-5.00%]) is comparable to overall *M. guttatus* diversity, and exceeds diversity within northern or southern *M. guttatus* collections (Figure 1D). We derive a simple estimate of split time between *M. guttatus* and *M. nasutus*, under a model assuming no introgression, as we did above to estimate the split between focal northern and southern *M. guttatus* samples. Using the difference between divergence of *M. nasutus* to the southern, allopatric *M. guttatus* sample (to minimize the influence of recent introgression and historical divergence between *M. guttatus*' genetic clusters) and a proxy for diversity in an ancestral population (southern *M. guttatus*), we estimate that *M. nasutus* and *M. guttatus* split approximately 200 ky ago ($\tau$= ($\pi_{S\ Nas\ x\ AlloSouthGut}$ - $\pi_{S\ SouthGut}$) / 2$\mu$ = 196 ky [181 ky - 212 ky]).

As a complementary inference of historical patterns of divergence within *M. guttatus* and between species, we applied Li and Durbin's implementation of the pairwise sequentially Markovian coalescent (PSMC) (32) to pairwise combinations of focal haploid genomes (Figures 1E and Figure S1-S5). Since we sample from a structured population, the inferred large recent population sizes likely represent reduced coalescent rates caused by population structure, rather than dramatic recent increases in $N_e$. Thus, the PSMC inference of larger recent population sizes between northern and southern *M. guttatus* (SLP x AHQT) compared to within these groups (SLP x DPRG and CACG x AHQT, Figures 1E and S1) likely reflects the strong genetic structure within *M. guttatus*. Similarly, the large recent population sizes inferred within both northern and southern samples likely reflect substructure within northern and southern regions.

We also use this PSMC analysis to roughly estimate a split date, by assessing when the inferred coalescent rate between species decreases (*i.e.,* the population size estimate increases) relative to the rate within *M. guttatus* (see 32). In doing so, we focus on the southern *M. guttatus* samples that fall closest to *M. nasutus* in our nj tree. The inferred coalescent rate between *M. nasutus* and southern *M. guttatus* (SLP x KOOT, gray line) decreases relative to the rate within southern *M. guttatus* (SLP x DPRG, dark blue/navy line), *i.e.*, the lines diverge, from ~500 to ~200 kya, suggesting either a gradual split between species over that time span, or a hard split sometime within that range (Figures 1E and S2). This result is qualitatively similar to our estimate based on synonymous nucleotide variation among these samples.

*Genomic consequences of the transition to selfing:*

We find that patterns of genomic variation within *M. nasutus* reflect the genomic consequences of a recent transition to selfing. Synonymous diversity within *M. nasutus* ($\pi_S$ = 1.09% [1.03%-1.14%], Figure 1D) is one fifth that observed within *M. guttatus*, consistent with a high rate of genetic drift since *M. nasutus*' origin. Moreover, most ancestral variation in *M. nasutus* has been homogenized: of the fixed differences between *M. nasutus* and *M. guttatus*, 90% are derived in *M. nasutus* and 10% are derived in *M. guttatus* (when polarizing by *M. dentilobus*). Although *M. nasutus* has lost much of its ancestral variation, shared variants still constitute a much higher proportion of its polymorphism (50%) relative to an equally sized sample of *M. guttatus* (10%). This suggests *M. nasutus*' genetic diversity disproportionately comes from genomic regions retaining ancestral variation.

We find an excess of high-frequency derived synonymous mutations in *M. nasutus* (Figure S6), suggesting that its population contracted recently. By contrast, we observe slightly more rare synonymous alleles than expected under demographic and selective equilibrium in *M. guttatus*, likely reflecting the structure and/or history of introgression in these samples, or weak selection against unpreferred codons.

The distribution of synonymous diversity in 5 kb windows across the genome (Figure 2A) bolsters the view that *M. nasutus*' genomic diversity is a mixture of closely related genomic regions that rapidly coalesce in the small *M. nasutus* population, and distantly related regions that do not coalesce until joining a large *M. guttatus*-like ancestral population. In pairwise comparisons of sequence diversity within *M. nasutus*, half of the genomic windows are differentiated by $\pi_S$ < 0.5% (corresponding to ~170 thousand years of divergence), reflecting recent common ancestry since the species split. On the other hand, one third of such windows are differentiated by $\pi_S$ > 2.0%, reflecting deep ancestry in a large ancestral population (Figure 2A).

These findings contrast sharply with comparisons within *M. guttatus*, as well as between *M. nasutus* and allopatric *M. guttatus* samples, for which recent common ancestry since the species split is rare ($\pi_S$ < 0.5% for less than 1.5% of 5kb windows) and deep coalescence is the norm (mode $\pi_S$ = 4%, Figure 2A). Under the neutral coalescent, a pair of lineages will fail to find a common ancestor with each other by generation $\tau$ with probability $e^{-t/N_{e^*}}$, where $N_{e^*}$ is the (constant) effective number of chromosomes. Therefore, the observation that half of our windows share a common ancestor in the past ~170 ky, by an admittedly crude calculation, predicts a population size between 150k and 250k effective chromosomes (compared to estimated $N_{e*}$ of 1.25 in *M. guttatus* from synonymous diversity, above). This tenfold reduction in effective population size as compared to *M. guttatus* far exceeds both the twofold decrease in $N_e$ expected to accompany the evolution of selfing and the fourfold decrease calculated by the difference in intraspecific variation.

Likewise, our PSMC results strongly support a history of extensive recent shared ancestry and the incomplete sorting of ancestral diversity in *M. nasutus*. We infer a dramatic decline in *M. nasutus'* effective population size after it split from *M. guttatus* (compare red and black-gray lines Figure 1E, see also Figure S3), suggesting that the evolution of selfing roughly coincided with *M. nasutus'* split from *M. guttatus*. We caution, however that interpretation of PSMC's estimated population size in *M. nasutus* is not straightforward. This is because the transition to selfing reduces the population recombination rate more than the population mutation rate (13); however, Li and Durbin's (32) implementation of the PSMC assumes that both these values change proportionally with the historical effective population size.

Across the genome, the mosaic nature of ancestry within *M. nasutus* is apparent as long contiguous regions of recent common ancestry (colored windows in Figures 2B and S7) interrupted by regions of deep ancestry, due to incomplete lineage sorting and/or historical introgression (white windows in Figure 2B and S7). This block-like ancestry structure results in extensive linkage disequilibrium (LD) in *M. nasutus*. In contrast to *M. guttatus*, for which the sample pairwise LD drops halfway towards its minimum values within only 15-20 base-pairs, LD in *M. nasutus* decays much more slowly, not dropping halfway towards its minimum values until 22 kb (Figure 2C). This represents a thousand-fold difference in the decay of LD, as compared to a more modest ten-fold reduction in the effective population size between *M. nasutus* and *M. guttatus*. This dramatic difference in the scale of LD between *Mimulus* species is likely due to a major reduction in the effective recombination rate within the selfing *M. nasutus*. Following Nordborg ((13) Eq 1), we use the comparison of the population scaled recombination and mutation rate to estimate an effective selfing rate of 99% in *M. nasutus*.

Patterns of sequence variation suggest a reduced efficacy of purifying selection in *M. nasutus*, a result consistent with extensive genetic drift and/or linked selection within *M. nasutus*. All *M. nasutus* samples contain more premature stop codons than any *M. guttatus* sample (*M. nasutus*: mean 124, range = 121-126, *M. guttatus*: mean 95.5, range = 86-102), and a large proportion of these premature stops are at high frequency in *M. nasutus* (Figure 2D). For 27 of the 29 fixed differences for a premature stop codon, *M. nasutus* carries the premature stop and *M. guttatus* carries the intact allele. Additionally, after standardizing by synonymous variation, we observe an excess of putatively deleterious, non-synonymous variation in *M. nasutus* relative to *M. guttatus* $\pi_N/\pi_S$ = 0.197 [0.192-0.203] and 0.157 [0.155-0.160], respectively). However, this difference is not yet reflected in divergence between the species ($d_N/d_S$ = 0.156 [0.154-0.159]), presumably because interspecific sequence differences largely reflect variation that predates the origin of selfing in *M. nasutus* rather than the relatively few mutations accrued within the past ~170 ky. This pattern of elevated $\pi_N/\pi_S$ in selfing species but only modest $d_N/d_S$ between selfers and their close relatives is common (14), even in genome-wide analyses (*e.g.,* (19)).

*Ongoing gene flow and its consequences:*

Somewhat surprisingly, given the placement of the sympatric, northern *M. guttatus* sample (CACG) far from *M. nasutus* in our neighbor-joining tree and PCA analyses (Figures 1B and 1C, respectively), CACG has the lowest level of nucleotide divergence to *M. nasutus* (Table S2). To test whether these seemingly contradictory observations are due to introgression, we used Treemix, an unsupervised method to construct a phylogeny featuring admixture events (33). Treemix finds a clear signal of admixture from *M. nasutus* into the population ancestral to CACG (Figure 3A). This result holds across a range of different sample subsets (Figure S8). Some Treemix analyses also suggest introgression from *M. nasutus* into the population ancestral to our southern, sympatric *M. guttatus* (DPRG); however, the manifestation of this second signal varies across sample subsets (Figure S8, see supplemental results).

Further quantitative evidence for introgression in the CACG sample comes from the bimodal distribution of divergence between *M. nasutus* and CACG in 5kb windows (dark purple lines in Figure 3B). Indeed, for most genomic windows, CACG shows deep divergence with *M. nasutus* (half of windows have more than 5% sequence divergence), but approximately one tenth of windows show nearly no divergence ($\pi_S$ < 0.5%). We find a similar, but subtler, bimodal distribution of divergence between *Mimulus* species featuring the sympatric southern *M. guttatus* sample (2.5% of windows comparing DPRG and *M. nasutus* exhibit less than 0.5% sequence divergence – dark blue lines in Figure 3B). We do not observe this pattern in the allopatric samples in which fewer than 0.5% of windows are less than 0.5% diverged from *M. nasutus* (light lines in Figure 3B and S7). The PSMC analysis also reflects ongoing introgression in sympatry, as it highlights recent common ancestry between sympatric *M. guttatus* samples (CACG and DPRG) and *M. nasutus* (Figures S4 and S5). Moreover, genomic regions of low interspecific divergence are spatially clustered (Figures 3C and S7), consistent with recent and ongoing introgression that is slowly broken down over generations of recombination.

We utilize this spatial distribution of windows of low interspecific divergence to probabilistically infer regions of recent *M. nasutus* or *M. guttatus* ancestry across the genomes of our four focal *M. guttatus* samples (Figure 3C) using a Hidden Markov Model ('HMM' see METHODS). From this HMM, we estimate recent *M. nasutus* ancestry for 15.1%, 5.7%, 1.1% and 0.6% of sympatric northern (CACG), sympatric southern (DPRG), allopatric southern (SLP), and allopatric northern (AHQT) *M. guttatus* genomes, respectively. The non-zero admixture proportions inferred in allopatric samples likely reflect low levels of admixture into allopatric *M. guttatus* and/or mis-assigned regions of incomplete lineage sorting.

To learn about the timing of admixture, we find contiguous regions of individual *M. guttatus* genomes with a greater than 95% posterior probability of *M.*

*nasutus* ancestry and display the length distribution of these blocks in Figure 3D. The mean admixture block lengths are 132 kb (~0.74 cM, n=227 blocks) and 18.6 kb (~0.10 cM, n=350 blocks) for northern (CACG) and southern (DRPG) sympatric *M. guttatus* samples, respectively. Because our HMM occasionally breaks up what seem to be contiguous blocks of admixed ancestry, we instituted a range of strategies to fuse admixture blocks, none of which greatly influenced our block length distributions (see supplementary results and Figure S9). Under a single temporal pulse of gene flow, admixture block lengths are exponentially distributed with a mean length (in Morgans) that is the reciprocal of the number of generations. With this model, we estimate 135 and 962 generations since admixture for CACG and DPRG samples, respectively (Table S5). Because these estimates are much more recent than the split time of around 200 kya, this pattern cannot be explained by incomplete lineage sorting.

While this pulse model provides an intuitive summary of time back until an admixed region finds itself in an *M. nasutus* sample, our data are inconsistent with a model of a single admixture pulse. Specifically, the block length distribution of *M. nasutus* ancestry in *M. guttatus* samples is too variable to be consistent with a single admixture time, and so these estimates should be viewed as average times back to an admixture event. Furthermore, these blocks come from far enough back in the past that every single block is likely derived from a separate admixture event, rather than being derived from a single *M. nasutus* parent (Table S5, see also Methods and supplementary results for quantitative support for these points).

We also detected a signal of introgression from *M. guttatus* into *M. nasutus*, despite the numerous difficulties presented by the short scale of LD in *M. guttatus* and the similarity between interspecific divergence and *M. guttatus* diversity. We overcame this challenge by reasoning that admixture from *M. guttatus* into *M. nasutus* would result in long genomic regions that are more closely related to nearby *M. guttatus* samples than to other *M. nasutus* samples. Therefore, we focus on long contiguous regions (> 20 kb) for which one 'outlier' *M. nasutus* sample does not coalesce with any other *M. nasutus* samples until before the species split (as inferred by pairwise $\pi_S$ > 1%, alternative thresholds explored in the supplement), and compare synonymous divergence of outlier individuals and non-outliers in non-overlapping 20 kb windows to northern and southern allopatric *M. guttatus* samples. By excluding sympatric *M. guttatus* samples, we avoid conflating introgression into *M. nasutus* with the extensive signature of introgression from the opposite direction (*i.e.*, *M. nasutus* into *M. guttatus*).

Pooling across northern *M. nasutus* samples (NHN, Koot, CACN), outlier windows are more often genetically closer to the northern *M. guttatus* sample (AHQT) than are the non-outlier windows (272 of 490, one sided binomial test against the null expectation of 50% P = 0.008). However, this result is individually significant only for our most geographically remote *M. nasutus* sample, NHN (For 141 of 289 outlier windows NHN are closer to northern *M. guttatus* than are the non-outlier *M. nasutus*). Additionally, the southern *M. nasutus* sample, DPRN had the smallest proportion of outlier windows resembling northern *M. guttatus*. By contrast, no samples have a disproportionate share of outliers genetically close to southern *M. guttatus*, SLP (Table S6), consistent with little or no introgression into DPRN (note, however, that this comparison is likely underpowered because of the genetic proximity of southern *M. guttatus* to the population that founded *M. guttatus*). In Table S6 and in the supplementary results, we show that there is significant support for introgression from *M. guttatus* into *M. nasutus* over a broad range of parameter choices for designating outlier windows – while no samples are genetically closer to southern *M. guttatus* in outlier windows more often than expected by chance, NHN individually and/or pooled northern *M. nasutus* samples are more often like northern *M. guttatus* than expected under incomplete lineage sorting.

*Divergence between M. nasutus and sympatric M. guttatus samples decreases with increasing local recombination rate:*

We observed a genome-wide negative relationship between absolute synonymous divergence (*i.e.*, the mean number of pairwise sequence differences at synonymous sites) and the local recombination rate in both sympatric *M. guttatus* samples (DPRG x Nas: Spearman's $\rho = -0.080$, P = 0.0008, CACG x Nas: Spearman's $\rho = -0.0718$, P = 0.0027). These sympatric *M. guttatus* samples display approximately a 5% reduction in synonymous nucleotide divergence in genomic regions with greater than average recombination rates. This signal is substantially weaker in the allopatric southern *M. guttatus* sample (SLP x Nas: Spearman's $\rho = -0.0521$, P = 0.0297), which is nestled within the range of *M. nasutus*, and altogether absent in comparisons with the allopatric northern *M. guttatus* sample, which occurs well outside of *M. nasutus'* range (AHQT x Nas: Spearman's $\rho = -0.0261$, P = 0.2768). This result holds after accounting for the potential confounding effect of sequencing depth and the influence of synonymous divergence to *M. dentilobus* (a proxy for mutation rate variation, *see* Methods, Supplementary results and Table S7).

      This negative relationship between divergence and the recombination rate in sympatry is consistent with selection against *M. nasutus* ancestry reducing effective gene flow at linked sites (see (34), for related theory and tentative evidence to date), and inconsistent with alternative scenarios. Specifically, neutral processes cannot generate this correlation because mean neutral coalescent times are independent of recombination rates (35). Selective sweeps and background selection acting simply on *M. guttatus* ancestry would not influence expected substitution rates at linked neutral sites (36), and so while they may influence relative measures of divergence (34, 37, 38), they will not influence absolute divergence. Moreover, since we observe no relationship between diversity and the recombination rate within *M. guttatus* (Spearman's $\rho = -0.0275$ P = 0.244) or *M. nasutus* (Spearman's $\rho = 0.0291$ P = 0.218), it seems unlikely that linked selection processes within species (background selection and selective sweeps) could explain this result.

      Finally, we note that genomic regions with the lowest recombination rates have the highest densities of centromeric and TE-like repeats, and in contrast, high-recombination regions have the highest local gene-densities per physical distance (data not shown). However, there is no obvious link between this observation and

the consistent negative correlation between divergence and recombination between *M. nasutus* and sympatric, but not allopatric, *M. guttatus*.

**Discussion**

*Speciation history and the origin of M. nasutus:*

Genetically, *M. nasutus* clusters with central Californian *M. guttatus* samples, suggesting that speciation post-dated the differentiation of some *M. guttatus* populations. Thus, speciation in this pair is best described as a `budding' of *M. nasutus* from *M. guttatus*, rather than a split of an ancestral species into two. We observe an approximate coincidence between the timing of divergence and the decline in population size in *M. nasutus* (as inferred from our PSMC analysis), likely as a result of the transition to selfing being linked to speciation (see (10) for phylogenetic evidence of this link in the Solanaceae and (39, 40) for a likely case in *Capsella*). Future genomic analyses across the *M. guttatus* complex and other species groups will facilitate an in-depth view of the causes and consequences of speciation by the budding of selfing and/or endemic populations off of a widespread parental species, and the commonness of this mode of speciation.

We estimate that *M. nasutus* split from a *M. guttatus* population within the last two hundred to five hundred thousand years (with our estimate of ~200 ky, inferred from differences in synonymous sequence differences within and between species, and the estimate of 500 ky corresponding to conservative estimates of population splits from the PSMC). This lies between the ~50 ky separating selfing *Capsella rubella* from outcrossing *C. grandiflora* (19, 41) and *Arabidopsis thaliana* which has potentially been selfing for over a million years ((42), having split from *A. lyrata* ~3-9 Mya (43)). Although 200 ky represents a relatively short time evolutionarily, it implies that *M. nasutus* managed to survive numerous dramatic bioclimatic fluctuations.

*The transition to selfing and its genomic consequences:*

The transition from outcrossing to self-fertilization in *M. nasutus* has had clear consequences on patterns of genomic variation. In *M. nasutus*, linkage disequilibrium exceeds that in *M. guttatus* by three orders of magnitude. This result suggests a high selfing rate in *M. nasutus* (estimated at 99%, above), consistent with direct estimates from field studies (23). We observe a four-fold drop in diversity and infer a ten-fold reduction in effective population size in *M. nasutus* compared to *M. guttatus*, values far exceeding the two-fold decrease in $N_e$ expected as a direct consequence of selfing (44, 45). This more than two-fold reduction in $N_e$ of selfing populations relative to their outcrossing relatives has been identified in other species pairs (16, 41), and may be partially due to extreme founding bottlenecks and/or frequent colonization events and demographic stochasticity that further increase the rate of genetic drift (12). Additionally, due to the lower effective recombination rate in selfing species (13), the effect of linked selection is magnified (46-48), further reducing diversity.

Selfing populations are expected to experience a reduced effectiveness of purifying selection accompanying the drop in effective population size and recombination rates (14, 48, 49). Consistent with these predictions, *M. nasutus* has accumulated numerous putatively deleterious mutations, including nonsynonymous variants and premature stop codons. Presumably, this elevation in radical genetic variants reflects a reduction in the efficacy of purifying selection due to a high rate of genetic drift and linked selection, as well as perhaps the escape of some genes (*e.g.*, loci involved in pollinator attraction) from the selective constraints they faced in an outcrossing population (*e.g.,* (16)).

*Selfing as a reproductive barrier and its significance for ongoing gene flow:*

Despite multiple reproductive isolating barriers, including mating system differences we find ongoing, bidirectional introgression between *M. guttatus* and *M. nasutus*.

Evidence of ongoing introgression from the selfer, *M. nasutus,* into the outcrosser, *M. guttatus,* is particularly stark. There are numerous evolutionary implications of introgression from selfers to outcrossers. Introgression of deleterious mutations accumulated in selfers may introduce a genetic load to outcrossers. This burden would result in selection against genetic material from selfers in hybridizing outcrossing populations, and could ultimately favor reinforcement of reproductive isolation. Alternatively, such introgression could provide a multi-locus suite of variation facilitating self-fertilization, and other correlated traits (*e.g.,* drought resistance and rapid development), in favorable environments, as appears to be the case in introgression between wild and domestic beets (*Beta vulgaris*, (50)).

Evidence of introgression from *M. guttatus* into *M. nasutus* is subtler, but is potentially critically important. Even relatively low levels of introgression into a selfer may rescue the population from a build up of deleterious alleles, and reintroduce adaptive variation, and so may lower its chances of extinction, a fate considered likely for most selfing lineages (51, 52). However, before potentially rescuing a selfing population from extinction, genomic regions introduced from outcrossing species must themselves survive a purging of deleterious recessive alleles.

Higher rates of introgression from *M. nasutus* to *M. guttatus* would be consistent with the prediction that backcrosses should be asymmetric – because bees preferentially visit plants with larger flowers (53, 54) and/or larger floral displays (55, 56), both features of *M. guttatus*, visits to *M. nasutus* and F1 hybrids are likely preceded and followed by visits to *M. guttatus* (23, 29). Consistent with this prediction, direct estimates of hybridization in the DPR sympatric population reveal that $F_1$ hybrids are the product of *M. nasutus* maternal and *M. guttatus* paternal parents, respectively (23). However, we caution that it is considerably more challenging to identify introgression into *M. nasutus* than into *M. guttatus*, as the similarity between interspecific divergence and diversity in *M. guttatus* makes historical admixture difficult to separate from the incomplete sorting of *M. nasutus*' ancestral variation. We further note that, although asymmetrical introgression from selfers to outcrossers has been detected in other systems (*Pitcairnia* (57), and potentially in *Geum* (58, 59)), the relative contribution of selfing vs. other isolating barriers and/or selection is unclear. Dense sampling of sympatric and allopatric populations of outcrossing species experiencing ongoing gene flow with selfing relatives will allow for tests of these hypotheses. Importantly, the number, location and length-distribution of admixture blocks identified from genomic analyses provide information about the longer-term consequences and pace of introgression between selfers and outcrossers.

*Selection against hybrids and implications for species maintenance:*

The numerous short blocks (in addition to long blocks) of *M. nasutus* ancestry observed in *M. guttatus* suggest that *M. nasutus* ancestry can potentially persist in an *M. guttatus* background for many generations. Despite this, *M. guttatus* and *M. nasutus* are still ecologically and genetically distinct. While this differentiation could arise via a balance of genetic drift and gene flow, there is a possible role for selection in the maintenance of species boundaries.

We identified a genome-wide signature of selection against introgression of *M. nasutus* ancestry in *M. guttatus*, in the form of a negative relationship between the local recombination rate and absolute divergence. This relationship was highly significant in both sympatric comparisons, but only weakly significant in parapatry, and insignificant in allopatry. Additionally, we did not find a relationship between recombination and diversity within either species. Moreover, unlike a negative relationship between the recombination rate and relative measures of differentiation, such as $F_{st}$ or the number of fixed differences (e.g., 60, 61, 62), this finding cannot also be explained by a high rate of hitchhiking or background selection within populations since the species split (34, 37, 38). Instead, it seems more consistent with *M. nasutus* ancestry being selected against more strongly in regions of low recombination due to linkage with maladaptive alleles that introgression would introduce (suggesting that the genome has congealed in low recombination regions, (63, 64)). Previously this result had been hinted at by higher absolute divergence near the breakpoints of inversions (*e.g.,* (65, 66)), or in centromeres relative to telomeres (67), but to our knowledge this is the first time this basic prediction has been demonstrated genome-wide.

Further work, including experiments measuring selection on genetic variants in the wild, and larger sample sizes from both allopatric and sympatric populations, is needed to pinpoint which (if any) genomic regions are particularly strongly selected against in hybrids. Genetically mapped loci for adaptive interspecific differences (68) and hybrid inviability and sterility (28) are promising candidates.

**Materials and Methods**

*Mimulus sampling and whole genome sequencing*

We utilized a combination of existing (downloaded from the NCBI SRA, sequenced by 69) and newly generated whole genome sequence data from 19 different lab and/or naturally inbred *Mimulus* accessions, including 13 *M. guttatus*, 5 *M. nasutus*, and 1 *M. dentilobus* individual as an outgroup. Samples varied in their geography and life history. Mean sequencing depths range from 2X to 25X, and read lengths include 36, 76, and 100 base pair paired end reads. We present SRA accession numbers as well as depth, read length and additional sample information in Table S1, and note that we obtained the DPRG sequence data directly from the U.S. Department of Energy Joint Genome Institute. Our analysis included newly generated whole genome sequences from five lines (CACG, CACN, DPRN, NHN, and KOOT), and we present details of sequence generation in the supplementary methods.

*Genome sequence alignment, SNP identification and annotation*

We aligned paired end reads to the *M. guttatus* v2.0 reference genome (69) using Burrows-Wheeler Aligner (bwa (70)) with a minimum alignment quality threshold of Q29 (filtering done using SAMtools (71)). Alignment-processing details can be found in the supplementary methods. We produced a high quality set of invariant sites and SNPs simultaneously for all lines using the GATK Unified Genotyper, with a site quality threshold of Q40 (72, 73). For all analyses described below, we exclusively used genotype calls from reference scaffolds 1-14, corresponding to the 14 chromosomes in the *Mimulus* genome. For all analyses (except PSMC, which requires a consistently high density of data, see below), we set also set a strict minimum depth cutoff of 10 reads per site (unless otherwise noted), removed triallelic sites, and censured genotypes at sites where individual depth was two standard deviations away from mean depth. To assign genotypes at heterozygous sites, we randomly selected one of two alternate alleles. Such heterozygous sites are not concentrated in long genomic regions and account for approximately 1% and 2% of synonymous SNPs in average focal *M. nasutus* and *M. guttatus* samples, respectively.

Following the filtering steps described above, all remaining genic genotype calls were classified as zero, two, three, or fourfold degenerate using the *Mimulus guttatus* v2.0 gene annotations provided by phytozome (69).

*Data analysis*

<u>NJ Tree and PCA</u>

We first analyzed broad patterns of genomic differentiation captured by principle component analysis and a neighbor-joining (nj) tree for all M. guttatus and M. nasutus samples, rooting our tree by the outgroup, *M. dentilobus*. For both these analyses (and the Treemix analyses) we downsampled to 1000 synonymous SNPs per chromosome with more than one copy of the minor allele. In this downsampling we insisted that no focal samples have missing genotypic data and then prioritized by the number of individuals with sequence data (see supplementary methods). After this downsampling, all individuals had genotypic data at more than 60% of SNPS, and 97% of these SNPs had genotypic data for at least 16 of our 19 individuals. We make the nj tree with the nj function in the R program (74), ape (75), and root it by the outgroup, *M. dentilobus*, We construct our PCA with customized R scripts accounting for missing genotype data described in the supplementary methods.

Nucleotide diversity and divergence

Preliminary analyses of nucleotide variation showed a strong influence of sequence features such as read depth and length (Table S3). In general, samples with low depth and short reads are less different from all other sample than are samples with high depth and long reads (Tables S3 and S4), a likely consequence of difficulties in aligning short reads when they differ substantially from the reference. We therefore focus all following analyses on eight samples with high and consistent mean read depths (13X-24X) and read lengths (all 100 bp paired end reads).

*Divergence and diversity:*
We quantified patterns of synonymous and nonsynonymous sequence variation at fourfold and zerofold degenerate sites, respectively. For each pairwise comparison, we count the number of pairwise sequence differences and number of sites for which both samples have data above our quality and depth thresholds. To generate confidence intervals for our point estimates of diversity and divergence that acknowledges the non-independence of sequence variants due to linkage disequilibrium, we resample 100 kb windows with replacement.

*Allele Frequency Spectrum (AFS):*
To polarize the AFS we examined all sites passing our depth and quality thresholds for *M. dentilobus* as well as all focal samples. For sites polymorphic in of focal samples, we labeled the *M. dentilobus* allele as ancestral.

*Premature stop codon identification:*
We searched for premature stop codons in each *Mimulus* accession using the *Mimulus guttatus* v2.0 gene annotations. We defined a mutant stop codon to be premature only if all three nucleotide sites were available for the codon, if it occurred in a gene for which at least 25% of the codons were available in the sample, and if the codon did not occur in the last 5% of all codons on the 3' end of the gene.

*Correlations between recombination rate and diversity (or divergence):*

We estimated genetic distances in centiMorgans (cM) using information from three existing *Mimulus* genetic linkage maps: one intra-population *M. guttatus* composite map (IMxIM), integrated from three different $F_2$ maps between individuals originating from the IM population (76), and two inter-specific $F_2$ maps between IM62 *M. guttatus* (the reference line) and SF *M. nasutus*, IMxSF_2006 (68) and IMxSF_2009 (C. Wu and J. Willis unpublished). All three linkage maps are available at http://www.mimulusevolution.org/. See supplementary methods for more details.

With this integrated map, we estimated a local recombination rate for every 100 kb window, smoothed by the mean rate in the surrounding 500 kb. In each window we calculated mean pairwise sequence diversity at synonymous sites and used a non-parametric spearman rank sign test to evaluate the relationship between synonymous sequence diversity and the local recombination rate. We excluded windows with fewer than 100 pairwise comparisons, and regions without recombination estimates. The final number of 100 kb windows included in each pair-wise comparison ranged from 1,773-2,023 (~177.3-202.3 Mb), or 60.5-69.0% of the reference genome. Moreover, the set of windows in each analysis largely overlapped - 1,756 windows were common to all analyses meaning that for a given test 87% to 99% of the windows were used in all other analyses. In the supplementary results and Table S7, we show that divergence to *M. dentilobus* and local depth do not influence our qualitative conclusions.

*PSMC:*

As a complementary inference of historical demography and differentiation, we applied Li and Durbin's implementation of the pairwise sequentially Markovian coalescent (PSMC) (32) to pairwise combinations of focal haploid genomes. See the supplementary methods for additional details. Due to high diversity in our dataset, we used a window size of 10 bp for PSMC analysis. For all comparisons, we ran PSMC for 20 iterations and used the following input settings: recombination/mutation ratio (r) = 1.25, Tmax = 10, number of intervals (n) = 60, number of population size parameters = 24, parameter distribution pattern = '1*4+1*4+1*3+18*2+1*3+1*4+1*6'. We represented time using a generation time of 1 year and a mutation rate of $1.5 \times 10^{-8}$. We note that choosing a fixed value for the recombination/mutation ratio is appropriate for comparisons between species and within *M. guttatus*; however, this does not capture the change in this ratio accompanying the transition to selfing in *M. nasutus*. Therefore quantitative estimates of population decline through time within *M. nasutus* are best viewed as very rough approximations. To generate a measure of variability in the PSMC estimates, we ran 100 bootstrap analyses for each pairwise comparison. See the supplementary methods for details.

Introgression Analyses

*Treemix:*
    As a test for recent admixture between *M. guttatus* and *M. nasutus*, we used Pickrell and Pritchard's (2012) Treemix method to model the evolutionary history of a group as a series of splits and gene flow events. For these analyses, we used the same subsample of 14,000 SNPs used for the nj tree and PCA described above. We specified a Treemix block size of 200 SNPs and estimated the evolutionary history including 1, 2, 3 or 4 admixture events. We also ran analyses with and without the reference line IM62 included, and with all nineteen *Mimulus* samples (see supplementary methods and results, as well as Figure S8).

*HMM:*
    We implement the forward-backward algorithm and posterior decoding as described by Durbin et al. (77) in a customized R script controlling for underflow (available upon request) to calculate posterior probabilities of *M. nasutus* or *M. guttatus* ancestry across all four focal *M. guttatus* samples.

    We take as our emissions the minimum pairwise π between our focal *M. guttatus* sample and all *M. nasutus* samples in non-overlapping 1 kb windows ($\pi_{close\_nas}$). We compared $\pi_{close\_nas}$ to the genome-wide distribution of π between a *M. nasutus* sample and its genetically closest *M. nasutus* sample and π between *M. guttatus* samples and the genetically closest *M. nasutus* sample, to calculate the emission probability of admixed or pure *M. guttatus* ancestry, respectively. In doing so we accounted for both the heterogeneity in the number of informative sites across the genome and the fact that we compare each *M. nasutus* sample to three others, while we compare each *M. guttatus* sample to all four *M. nasutus* samples (see supplementary methods).

    We calculate $t_{i,j}$, the probability of transitioning from ancestry of species *i* to ancestry from species *j* from α, the admixture proportion, and r, the product of the recombination rate per 1kb multiplied by a point estimate of the number of generations since admixture. We set $t_{i,j}$ to $t_{gut,gut} = (1 - r) + r (1 - \alpha)$, $t_{gut,nas} = r \alpha$, $t_{nas,gut} = r (1 - \alpha)$, and $t_{nas,nas} = (1 - r) + r \alpha$, and optimize these parameters with the Nelder-Mead algorithm implemented in the R function, optim, calculating the likelihood of our data given these parameters from the forward algorithm. α estimated in this way is similar to estimates from the proportion of low divergence windows presented in the text, suggesting that our data provide information about both α and r. We assume that r is constant across windows, ignoring the influence of the recombination rate on the transition rate.

*Inference of introgression history:*

The number of admixture blocks and our point estimate of admixture timing are strong evidence that admixture is not the result of a single chance *M. nasutus* ancestor in the history of these samples. To see this, consider that our current day genome is expected to be broken into $X$ chunks $v$ generations ago, where $X$ is the number of recombination events (*i.e*, the map length in Morgans (approximately 14.7 Morgans in *Mimulus*) times the number of generations, $v$), plus the number of chromosomes (14 in *Mimulus*). So, for example, the CACG genome is broken into $X$ = 2219 chunks at our point estimate of its admixture time, $v$ =150 generations ago (150 generations times 14.7 Morgans plus 14 chromosomes). These $X$ chunks are drawn from $2^v$ genealogical ancestors spread across many ancestors, meaning that CACG is expected to inherit $2219/2^{150} \sim 1/2^{139}$ chromosomal segments from a typical genealogical ancestor, and therefore under a point model of admixture the odds of CACG inheriting two ancestry blocks from the same ancestor is vanishingly small. Therefore, each of the 227 *M. nasutus* ancestry blocks observed in CACG descends from a different admixture event, implying a vast number of admixed ancestors in the history of the sympatric populations *M. guttatus*. The case is starker for DPRG, for which we infer a much older point estimate of the admixture time.

    Because recombination is a Poisson process, under a single admixture time, $v$, we expect the variance in admixture block length to equal the square of the mean block length. However the variance in admixture block lengths is inconsistent with this expectation for both southern (DPRG, mean = 0.0010 M, σ = 0.0013 M, Bootstrap P < 0.001), and northern (CACG, mean = 0.0074 M, σ = 0.0102M, Bootstrap P < 0.001) sympatric *M. guttatus* samples. This argues against a point model of admixture and suggests ongoing and sustained gene flow into *M. guttatus* where it is sympatric with *M. nasutus*. We acknowledge that these calculations are somewhat crude, especially because we assume a constant recombination rate genome-wide. Nonetheless, the extreme variation in admixture block length and the large number of blocks visually obvious in Figure S9 supports this qualitative result.

    We note that since our inference of the extreme improbability that two admixture blocks are derived from the same introgression event relied on a point model, we must soften this conclusion. It is plausible that a few young ancestry blocks in CACG are descended from the same admixture event, however, the rejection of a point admixture model strengthens our major conclusion that gene flow is ongoing and that our samples do not represent single admixture events.

*Inference of Introgression from M. guttatus to M. nasutus.*

To test for introgression from *M. guttatus* into *M. nasutus*, we took advantage of the structure of genetic variation in *M. nasutus* – for most of the genome all individuals are remarkably similar, and when this is not the case, one individual often differs sharply from all others. The genetic variation in such genomic regions has either been maintained from the stock of ancestral variation, or it has been reintroduced by introgression upon secondary contact. The extreme genetic variation and miniscule extent of linkage disequilibrium within *M. guttatus* makes these two alternative hypotheses nearly indistinguishable in any given region; however, by collating information across all such regions we can test the introgression hypothesis.

To do so, we find long regions (20 kb) of the *M. nasutus* genome for which one individual is a genetic outlier, as described above. In these regions, we ask whether the outlier is more closely related to a specific *M. guttatus* sample than are the non-outliers. Under incomplete lineage sorting, there is a 50% probability that this is the case. However, under an admixture model this probability is greater than 50% if the *M. guttatus* sample is more closely related to the potential admixture source than it is to the population that founded *M. nasutus*. See supplementary results for more details, and Table S5 for the robustness of this inference to our exact rules for identifying outlier windows.

We perform this test for all *M. nasutus* samples individually against two focal allopatric *M. guttatus* samples from the northern (AHQT) and southern (SLP) groups. In addition to testing the one sided hypothesis of whether the outlier is sample is more often like a give M. guttatus sample we also pool our northern M. nasutus samples to amplify any signal.

**Acknowledgements**

We thank Dan Rokhsar, Kerrie Barry, and others at the U.S. Department of Energy Joint Genome Institute who provided primary genome sequencing and support. We also thank the Willis lab at Duke University (especially Jennifer Modliszewski) for DNA extractions. This study was supported in part by resources and technical expertise from the Georgia Advanced Computing Resource Center (GACRC), a partnership between the University of Georgia's Office of the Vice President for Research and Office of the Vice President for Information Technology. We particularly thank Shan-Ho Tsai, Curtis E. Combs Jr., and Yecheng Huang at the GACRC for their technical expertise and/or support. Saravanaraj 'Raj' Ayyampalayam at the Quantitative Biology Consulting Group in the University of Georgia Institute of Bioinformatics provided technical advice. John Willis and Stephen Wright provided thoughtful comments, which significantly improved the quality of our manuscript. Funding was provided by the University of Georgia Research Foundation to Andrea Sweigart and by the Alfred P. Sloan Foundation to Graham Coop. The work conducted by the Joint Genome Institute is supported by the Office of Science of the U.S. Department of Energy under Contract No. DE-AC02-05CH11231.

*Figure 1: Divergence in Mimulus **A)*** A map of all samples with identity denoted by color. Numbered positions represent high depth, high quality 'focal samples.' One through four are *M. nasutus* samples NHN (1), Koot (2), CACN (3), DPRN (4), and five through eight are *M. guttatus* samples AHQT (5), CACG (6), DPRG (7) and SLP (8), respectively (see table S1). ***B)*** A neighbor-joining tree for all samples rooted by *M. dentilobous*. Colors uniquely distinguish samples with reds representing *M. nasutus* and blues and purples representing the southern and northern *M. guttatus* clades, respectively. The tree was constructed from the pairwise pairwise distance matrix described in the main text with the *nj* function in the R package, *ape* and smoothed with the function, *chronopl* with λ = 1, an implementation of Sanderson's nonparametric rate smoothing program, r8s (Sanderson 2002). ***C)*** A principle component analysis of these data, excluding *M. dentilobus*. ***D)*** The mean number of pairwise sequence differences per fourfold degenerate site ($\pi_S$) within and among *Mimulus* species and populations, including uncertainty via a block bootstrap. ***E)*** Demographic history as inferred by the PSMC. Inferred population size through time is shown for pairwise combinations of haploid genomes of *M. guttatus* and/or *M. nasutus* individuals. Black/gray = interspecific comparisons with allopatric *M. guttatus*. Blue and violet = intraspecific *M. guttatus* comparisons. Red = intraspecific *M. nasutus*. For each pair-wise comparison, the thick dark line represents the point inference and each lighter-colored, thin line represents 1 of 100 bootstraps (see supplementary methods).

***Figure 2:*** *Genomic consequences of the transition to selfing.* ***A)*** A histogram of pairwise sequence diversity ($\pi_S$) within and between species in overlapping 5kb windows. For interspecific comparisons we focus only on allopatric *M. guttatus* populations. The dotted line denotes $\pi_S$ < 0.5% or 170 ky of divergence. ***B)*** Moving along a part of chromosome four, for all M. nasutus samples, we color genomic regions in which the focal individual (y-label) and another *M. nasutus* sample, indicated by color, recently coalesce ($\pi_S$ ≤ 0.5%). White regions coalesce more distantly in the past ($\pi_S$ > 0.5%) and grey regions indicate insufficient density of informative sites. Tick marks on the x-axis indicate 1 megabase. ***C)*** Linkage disequilibrium (measured as $r^2$) within *M. nasutus* (red) and *M. guttatus* (blue), as a function of physical distance. ***D)*** The number of premature stop codons observed in one, two, three, or four *M. nasutus* (red) and *M. guttatus* (blue) samples.

***Figure 3:*** *Introgression of M. nasutus material into M. guttatus.* ***A)*** Treemix suggests introgression from *M. nasutus* in to sympatric *M. guttatus* samples. ***B)*** A histogram of interspecific pairwise sequence divergence by *M. guttatus* sample. ***C)*** Introgression across a chromosome - Moving along a part of chromosome four for all *M. guttatus* samples, we color genomic regions in which the focal individual (y-label) and a *M. nasutus* sample, indicated by color, recently coalesce ($\pi_S$ ≤ 0.5%). White regions coalesce more distantly in the past ($\pi_S$ > 0.5%) and grey regions indicate insufficient density of informative sites. Tick marks on the x-axis indicate 1 megabase. Purple bars above each focal individual denote greater than a 95% posterior probability of *M. nasutus* ancestry as inferred from our HMM. ***D)*** Admixture block length

distribution - The number of admixed blocks, as inferred by a greater than 95% posterior probability of *M. nasutus* ancestry from our HMM, longer than x.

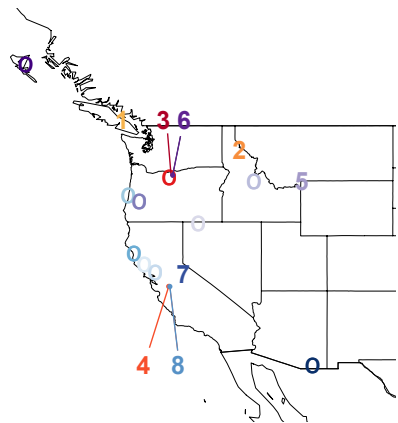
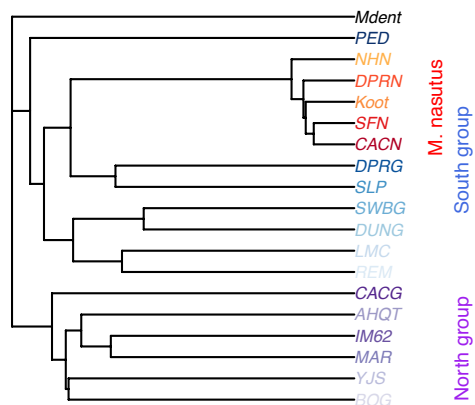
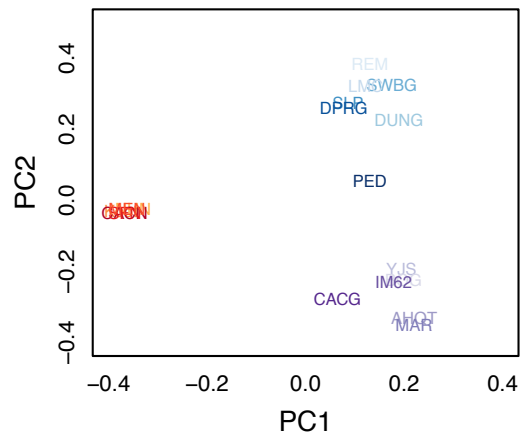
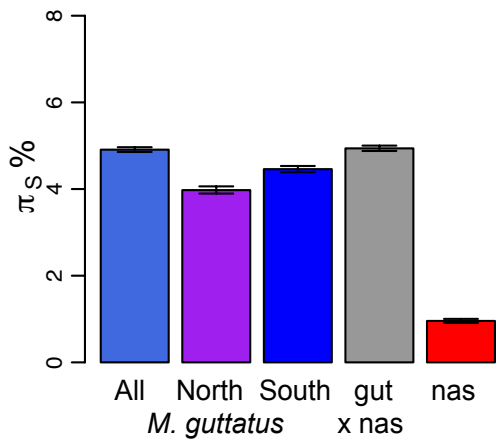
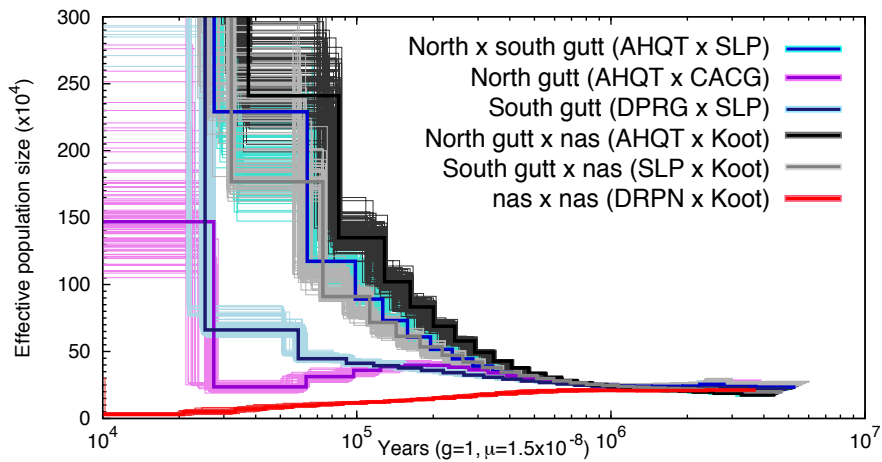

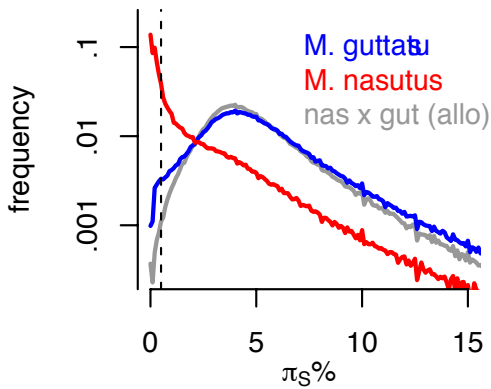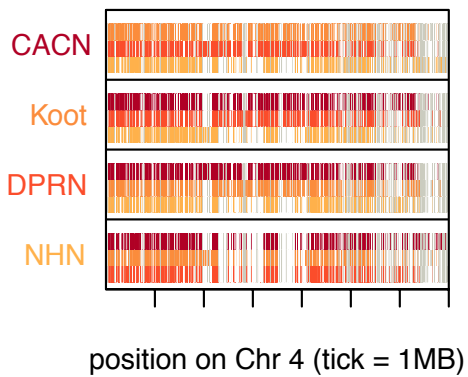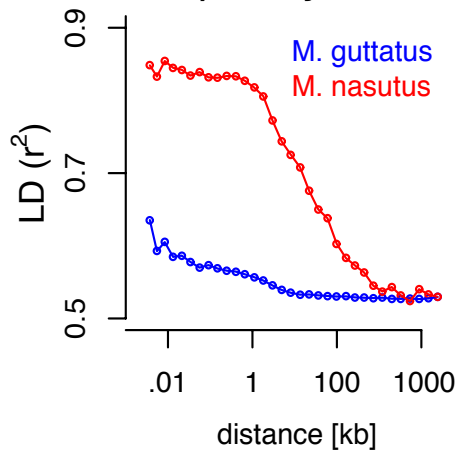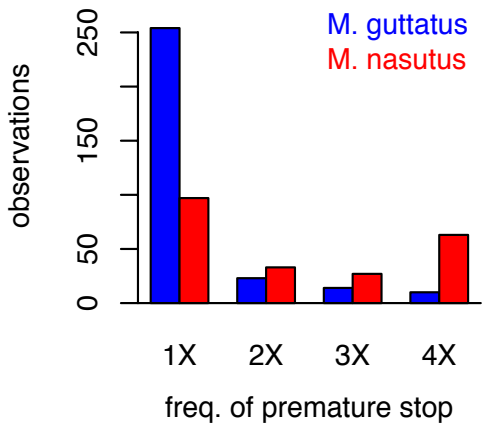

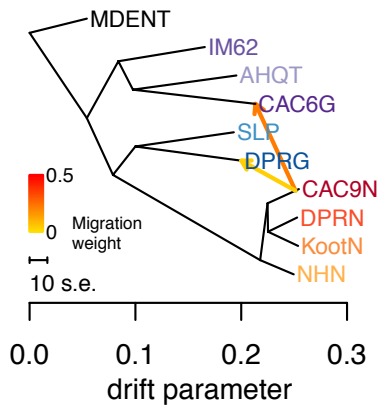
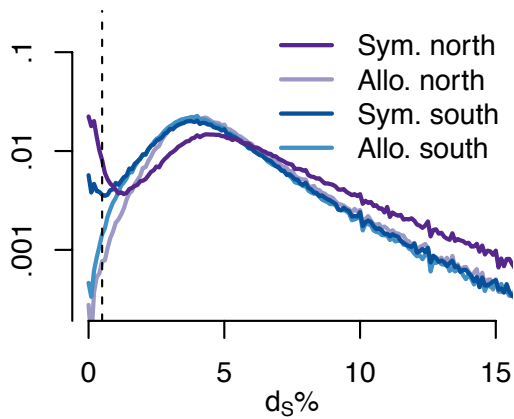
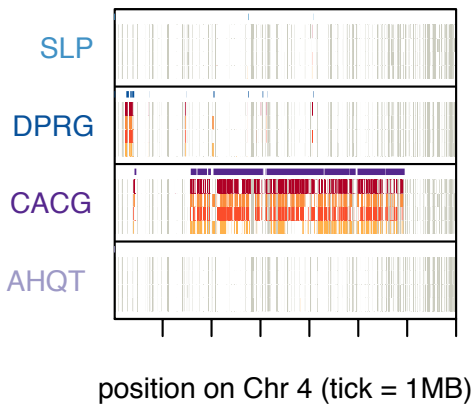
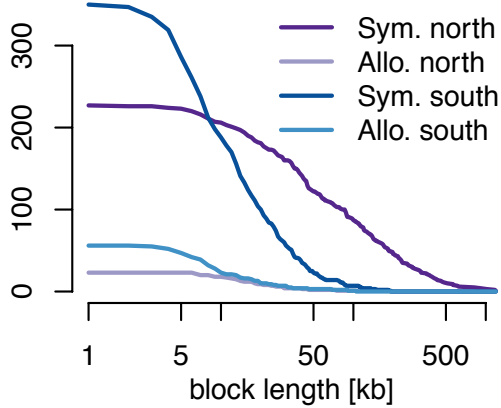

## Supplementary Material

**Supplementary methods**

*Sequencing:*

We generated new whole genome sequences for one *M. guttatus* (CACG) and four *M. nasutus* samples (CACN, DPRN, NHN, and KOOT). For these five samples, colleagues at Duke University extracted genomic DNA using a modified CTAB protocol (Kelly and Willis 1998) and RNAse A treatment. Sequencing libraries were prepared at the Duke Institute for Genome Sciences and Policy (IGSP) using standard Illumina Tru-Seq DNA library preparation kits and protocols, and sequenced on the Illumina Hi-Seq 2000 platform at the IGSP. Before sequence analysis of all samples, we removed potential contamination of sequencing adapters and primers with Trimmomatic (Lohse, Bolger et al. 2012) and confirmed removal using FastQC (www.bioinformatics.babraham.ac.uk/projects/fastqc/).

*Alignment processing*

After alignment, we removed potential pcr and optical duplicates using Picard (http://picard.sourceforge.net/). We did not filter reads with improper alignment flags (≤ ~5% of the mapped reads), however this had little effect on genotype calls (average proportion sequence difference between filtered and unfiltered datasets = $6.2 \times 10^{-5}$ (± $1.3 \times 10^{-5}$ SE) for five *Mimulus* lines varying in sequencing depth and read length). To minimize SNP errors around insertion/deletion polymorphisms, we performed local realignment for each sample using the Genome Analysis Tool Kit (GATK; McKenna, Hanna et al. 2010; DePristo, Banks et al. 2011).

*Downsampling for nj tree and PCA analyses:*

We use all nineteen samples for a genome-wide SNP analysis to learn about the genetic relationships and major components of genetic variation in these samples. For these analyses, we sampled 1000 fourfold degenerate SNPs per chromosome (14,000 in total), which had at least two copies of the minor allele, and prioritized SNPs by the number of samples with available genotype data. For all loci we had sample data for at least 14 of 19 samples, and for 97% of loci, we had data for at least 16 of 19 samples. Coverage across these 14,000 SNPs ranged from 60% to 100% per sample.

*PCA:*

We constructed a covariance matrix across pairs of individuals. To do so we evaluated the mean genotypic covariance across all 1000 focal sites for a pair of samples. We calculated principle components as the eigenvectors of this matrix. Customized R scripts for this operation are available from the authors upon request.

*PSMC input file generation and bootstrap analyses:*

To create pseudo-diploid genomes for PSMC analyses (Li and Durbin 2011), we first called the consensus sequence for each of our lines by running SAMtools mpileup (Li, Handsaker et al. 2009) on the final, locally realigned bam file for each line. Due to differences in overall coverage among chromosomes, we set the minimum coverage to 5X for chromosomes 1, 2, 4, 6, 8, 10, 13 and 14, and 1X for chromosomes 3, 5, 7, 9, 11 and 12. For each line, we set the maximum coverage for all chromosomes to 2 times the standard deviation plus the mean. We merged consensus sequences using Heng Li's seqtk toolset (https://github.com/lh3/seqtk), with a quality threshold of 20. For any site with residual heterozygosity, we randomly chose one allele.

To generate a measure of variability in the PSMC estimates of *M. guttatus* diversity and species divergence through time we ran 100 bootstrap analyses for each pairwise comparison. We used the PSMC utility splitfa to break up each pseudo-diploid genome into non-overlapping, similarly sized segments (resulting in 59 segments). To perform bootstrap analyses, we ran 100 separate PSMC analyses using the segmented genome as input and the –b (bootstrap) option. The bootstrap option randomly resamples with replacement from all of the segments to generate a unique/bootstrapped genome, similar in size to the original, and then runs PSMC on the bootstrapped genome. Note that bootstrapped genome sets were independent among different pairwise *Mimulus* comparisons. For our analyses we present both the point estimate using the full pseudo-diploid genome for each pairwise comparison (dark, thick lines) and the 100 bootstrap analyses (lighter, thin lines).

*Treemix analyses:*
Genotypes for these analyses consisted of the 14,000 biallelic SNPs used in our neighbor-joining and PCA analyses (above). We considered each line a population, with population allele counts being represented as '2,0' or '0,2' for the alternate genotypes, and '0,0' as missing data.

*HMM to identify introgression in M. guttatus:*
To make appropriate emission probabilities for our HMM we need to generate a comparable distribution of pairwise comparisons within our four *M. nasutus* samples and between focal *M. guttatus* samples and the four *M. nasutus* samples. We also must acknowledge the heterogeneity in the density of called sites (*i.e.*, sites where both samples surpass our quality cutoffs) across the genome and across individuals. To even out sample size (because each *M. nasutus* could be compared to three other *M. nasutus* samples, while each *M. guttatus* sample could be compared to four *M. nasutus* samples), we alternately left out one *M. nasutus* sample in our calculation of π between an *M. guttatus* sample and the nearest *M. nasutus*. We combined all values across the 16 classes of comparisons (the product of four *M. guttatus* samples and the four ways to leave out one *M. nasutus* sample) to calculate the empirical distribution of π to the nearest *M. nasutus* sample.

To accommodate the heterogeneity in the number of called sites, we bin all pairwise comparisons in 1kb windows by the number of sites with data for both lines (greater than the smaller bounds and less than or equal to the larger 0,5,10,20,50,75,100,250). Within each window, there are 3 pairwise comparisons. Among these, we select the comparison with the lowest pairwise π that is also in the bin with the most sites. In practice, this usually amounts to selecting the lowest pairwise π in a window, because in 65% of windows all three pairwise comparisons to a focal individual are in the same bin. Nonetheless, we make note of the number of pairwise comparisons for each minimum π value and use this as a second layer of conditioning, below. For each set of conditioning we calculate the frequency of windows with π in given discretized bins.

With our distribution of π to the nearest *M. nasutus* samples in hand, we can now calculate emission probabilities. We do so for each 1kb window, conditional on the largest bin of called sites and the number of pairwise comparisons with this number of called sites. For a given *M. guttatus* sample, we systematically leave out one *M. nasutus* sample, looping over each *M. nasutus* sample. We then find the emission probabilities for *M. guttatus* or *M. nasutus* ancestry by finding the proportion of appropriately binned minimum π values in our within *M. nasutus* comparisons and the proportion of minimum π values in *M. guttatus* to *M. nasutus* comparisons, respectively. Finally, we average these emission probabilities across the four ways in which we left out a *M. nasutus* sample.

*Recombination map:*

In order to approximate the genetic distance per physical unit (cM/kb) of the IM62 *M. guttatus* v2.0 reference genome (Joint Genome Institute), we accessed the map resources available at http://www.mimulusevolution.org. We began with the IMxIM map as an initial map because it contains linkage information from multiple individuals from the IM population. The IMxIM map also has the greatest number of mapped markers. To increase marker density we added markers from the two additional IMxSF maps not included in the IMxIM map. If flanking markers were shared between maps and if marker order was consistent, we assigned these additional markers a proportional genetic position in the IMxIM map according to their original recombination distance. We excluded entire regions where the genetic order of markers disagrees with the physical order of the reference genome, as well as regions distal to the first and last mapped marker on each chromosome; we did not estimate recombination in these regions or include them in our analyses with divergence. This conservative approach resulted in a final integrated map containing 285 markers with a total map length of ~14.7 Morgans (with genetic distances for ~256.5 Mb (87.5%) of the reference genome).

*Recombination rate and diversity (divergence):*

While calculating mean synonymous diversity in a window, we also calculate mean depth at synonymous sites and mean synonymous divergence to the outgroup, *M. dentilobus*. We then examine the spearman rank correlation of the local recombination rate and the residuals of the linear model where diversity is a function of divergence to *M. dentilobus* and mean depth at synonymous sites (Table S7).

**ANALYSES AND RESULTS**

*Pairwise comparisons:*
Values of $\pi_S$ and $\pi_N/\pi_S$ for each pairwise comparison are presented in Table S2. The mean number of pairwise sequence differences between *M. nasutus* and each focal *M. guttatus* sample is 4.54% (Northern sympatric – CACG), 4.76% (Southern sympatric – DPRG), 5.05% (Southern allopatric – SLP), 5.41% (Northern allopatric – AHQT).

To highlight the influence of read length and depth on estimates of diversity, in Table S3 we present mean $\pi_S$ and $\pi_N/\pi_S$ values across all population comparisons split by the number focal samples in a comparison (*i.e.*, zero means that neither of the samples is included in our detailed genomic analyses due to low depth or short reads). Note that for a given comparison between populations, diversity between two focal samples is much higher than that between two non-focal samples illustrating the influence of sequencing effort on diversity estimates. To avoid these effects we focus on our focal samples when discussing levels of diversity. We also note that we did not present in-depth genomic analyses of comparisons including the reference, IM62, because of unknown biases it may introduce.

We present values of pairwise $\pi_S$ between each focal sample with (above the main diagonal) and without (below the main diagonal) putatively introgressed regions (as inferred by a >95% posterior probability of *M. nasutus* ancestry in a genomic region of an *M. guttatus* sample) in Table S4. Reassuringly, after removing such regions, our two northern focal *M. guttatus* samples no longer differ in the number of pairwise sequence difference to *M. nasutus*, suggesting that our HMM performed very well for CACG (compare CACG and AHQT to *M. nasutus* samples above and below the main diagonal). Removing regions of inferred recent introgression in DPRG also increased its distance from *M. nasutus* samples; however, this sample is still genetically closer to *M. nasutus* than is the allopatric southern sample, SLP. This suggests that our HMM may have missed short (*i.e.* old) regions of introgression into DPRG and/or that even without introgression, DPRG is more closely related to the *M. nasutus* progenitor population than is SLP.

*Additional PSMC results:*
We present the results of additional PSMC analyses, including the effects of admixture on shared variation between *M. guttatus* and *M. nasutus*, and *M. nasutus*' population size decline, as well as 'zoomed-in' and 'zoomed-out' views of alternative figures (changing the y-axis limits) for each analysis. In Figure S1, we present a 'zoomed-out' view of Figure 1E providing a view of historical population splits and population size changes over time. The extreme variation in recent population sizes demonstrates both the effect of population structure within *M. guttatus* on population size estimates, and the lower accuracy of PSMC time estimates in recent history (Li and Durbin 2011). In Figure S2, we present a 'zoomed-in' view of the split between *M. nasutus* and southern *M. guttatus*. The approximate split date of ~200-500 kya is visible by evaluating roughly when the southern *M. guttatus* x *M. nasutus* curve (SLP x KOOT, gray) diverges from the southern *M. guttatus* curve (SLP x DPRG, blue; see Li and Durbin 2011).

We infer the history of population size decline in *M. nasutus* by running PSMC on all pairwise comparisons of our four high-coverage *M. nasutus* lines. From Figure S3, we observe that *M. nasutus*' decline in effective population size was coincident with divergence between the species, indicating that it is plausible that the evolution of selfing was associated with speciation and the origin of *M. nasutus*. The extreme reduction in *M. nasutus*' effective population size relative to *M. guttatus* is also evident from these analyses.

Finally, our PSMC analyses demonstrate an effect of admixture on the inferred history of divergence. We observe a reduction in the between species effective population size between *M. nasutus* and sympatric *M. guttatus*, relative to that between *M. nasutus* and allopatric *M. guttatus* (Figure S4, and S5 for the full, zoomed-out view). In Figure S4, relatively recent (*i.e.*, between ~10 and 70 kya) effective population sizes between *M. nasutus* and sympatric *M. guttatus* are reduced relative to allopatric comparisons, and roughly in the range of or even lower than population sizes within southern and northern *M. guttatus*, further supporting a history of ongoing and recent introgression.

*Robustness of introgression results*

Treemix
We explored our Treemix analyses over a range of different sample subsets and numbers of admixture events:
    *(A)* Focal samples and the reference (IM62) rooted by the outgroup
    *(B)* Focal samples rooted by the outgroup
    *(C)* All *M. nasutus* and *M. guttatus* samples rooted by the outgroup.
For each set of samples, we allowed one, two, three, or four historical admixture events (Figure S8). Regardless of sample subset and the number of admixture events allowed, we always see strong evidence of introgression from *M. nasutus* into CACG, a result consistent with all analyses in this manuscript. However, the other clear signal of introgression observed in our genomic analyses – introgression from *M. nasutus* into DPRG, was only observed when we allow for more than one introgression event and analyze all focal samples and the reference genome (Figure S8 A.2-A.4). When we limit our analysis to focal samples (rooted by the outgroup) and allow for two or more introgression events, treemix places an introgression arrow from northern *M. guttatus* samples to SLP (Figure S8 B.2-B.4). We view this result as consistent with introgression of *M. nasutus* material into DPRG, as introgression of the northern *M. guttatus* samples to SLP leaves DPRG as closer to the *M. nasutus* samples than SLP.

Combined with our genome-wide analyses, we view this as good evidence of introgression from *M. nasutus* into DPRG.

Block length distributions

In the main text, we used the length distribution of admixture blocks to provide a detailed view of the recent history of introgression of *M. nasutus* ancestry into *M. guttatus*. While this summary of the data contains much information, our inference of this distribution is likely imperfect. In practice, we may break up long admixture blocks or we may mislabel short genomic regions with low divergence as short admixture blocks.

In practice, both problems could confuse our inference. Miscalling short unadmixed regions and breaking up long regions into numerous smaller ones will both push back our inferred admixture time. Additionally, introducing short, false positive blocks may mislead us into seeing a mix of old and new admixture events, when in practice there was a single recent pulse. A major claim of our manuscript is that admixed *M. guttatus* samples are not simply early-generation hybrids, but rather represent on ongoing history of introgression. We therefore wish to ensure that these potential challenges to characterizing the block length distribution do not mislead our inference.

We use two strategies to ensure the robustness of our results. First, we 'heal' admixture blocks within X = {0,20,50,100} kb of one another (Figure S9), to guard against breaking up few long admixture blocks into more short ones. We also use our allopatric and putatively 'pure' *M. guttatus* samples to empirically control for the false positive admixture blocks. To do so, we alternatively use the block length distribution of AHQT and SLP and remove the closest matched block lengths in our other samples (note that we remove use the AHQT block length distribution in an attempt to better characterize the introgression history of SLP as well). By factorially implementing these controls, we see that our inference of ongoing introgression of *M. nasutus* into sympatric *M. guttatus* populations is robust. In all controls, we observe more variation in admixture tract lengths than would be expected under a simple point admixture model. Moreover, while removing young blocks and creating longer blocks creates a more recent estimated admixture time, our most recent estimated admixture time in CACG is 37 generations ago, arguing against a single recent admixture event. Even if admixture occurred 37 generations ago into CACG, it is very unlikely that a block from a given event at that time would survive to the present – and therefore gene flow is likely (relatively) consistently ongoing (Table S5).

Robustness of inferred introgression from *M. guttatus* into *M. nasutus*

In the main text we described our strategy of using outlier windows – regions where one *M. nasutus* sample differed radically from all others to infer historical introgression from *M. guttatus* into *M. nasutus*. The identification of outlier windows required numerous decisions; here we investigate the robustness of the signal of introgression to these choices.

The first was the $\pi_S$ cutoff differentiating outlier and non-outlier regions. We chose three alternative values for this cutoff – 0.5% (roughly corresponding to the expected level of differentiation since the species split), 2.0% (roughly corresponding to expected levels of variation within an ancestral *M. guttatus* population), and 1.0% (representing a compromise between these values). Within a given cutoff, we identify 20 contiguous overlapping sliding windows (with a 1 kb slide) where one sample differs from all others by $\pi_S$ greater than this threshold, while the others are differentiated from one another by $\pi_S$ less than this threshold. Although we always insist on 20 contiguous windows (representing 20 kb), we vary the window size, allowing it to take the value of 5, 10 or 20 kb (noted by **L** in Table S5).

Regardless of exact thresholds, we always see evidence for either introgression into NHN and/or introgression into the pooled collection of northern *M. nasutus* samples (CACN, NHN, and Koot), in the form of too many outlier windows being too close to AHQT (Table S6). By contrast, no samples are closer to SLP in outlier regions more often than expected by chance. However, as noted in the main text, our inability to identify introgression from southern M. *guttatus* into southern *M. nasutus* is likely underpowered because SLP may be too similar to the population that founded *M. nasutus*.

The relationship between divergence and recombination rate is not driven by sequencing depth or mutation rate variation

In the main text, we report a strong negative relationship between the local recombination rate (in 100 kb windows, smoothed over 500 kb) and absolute divergence between *M. nasutus* and sympatric *M. guttatus* samples at synonymous sites. We control for the potential confounds of the mutation rate (measured as divergence to *M. dentilobus*) and/or sequencing depth (at synonymous sites) in Table S7. To do so, we find the nonparametric correlation (Spearman's $\rho$) between the recombination rate and residuals of predicted divergence given local depth and/or divergence to *M. dentilobus* (where predictions come from the best fit linear model).

**Supplementary table legends**

*Table S1)* Detailed information about the biology, geography, inbreeding and sequencing of each sample analyzed in this manuscript. Gray-shaded rows denote focal samples used in our primary analyses.

*Table S2)* Pairwise sequence comparisons between all samples. We report the identifiers of our samples, the mean number of pairwise sequence differences at fourfold degenerate sites ($\pi_S$), the ratio of diversity at fully constrained and fourfold degenerate sites ($\pi_N/\pi_S$), the type of comparison (with regard to the population and species sample), and the number of focal samples involved in the comparison.

*Table S3)* A summary of pairwise sequence diversity in comparisons between all samples. We split this summary by the number of focal samples and the class of population comparison, and present both the mean number of pairwise sequence differences at fourfold degenerate sites ($\pi_S$), and the ratio of diversity at fully constrained and fourfold degenerate sites ($\pi_N/\pi_S$).

*Table S4)* After removing regions of recent introgression, interspecific divergence corresponds to the topology of the neighbor-joining tree. We present the percent of fourfold degenerate sites that differ between samples before (above the main diagonal) and after (below the main diagonal) removing regions inferred to be recently introgressed. Note that while CACG is much closer to *M. nasutus* samples before removing introgressed regions than AHQT, these two samples are equally differentiated from *M. nasutus* after removing regions of introgression.

*Table S5)* Summary of the admixture block-length distribution and the robustness of our inference of admixture history. For each focal sample, we present (1) The inferred number of blocks of *M. nasutus* ancestry (2) The total length (in kb) of *M. nasutus* ancestry inferred in the focal sample, (3) The mean block length in centiMorgans = mean_block_length[bp] * total_map_length[cM] / genome_size[bp], where total_map_length ~ 1470 cM and total_genome_size ~ 2.6*10$^8$. (4) The standard deviation in block length (in cM). (5) ***T***, the expected number of generations since admixture = 1/mean_length(cM). (6) P(exp) – A one sided test of the hypothesis that our block length distribution is not more variable than expected under one admixture pulse at time ***T***. We obtained this probability by resampling blocks with replacement 1000 times and finding the proportion of resampling experiments with mean block lengths greater than the variance in block lengths. (7) The probability that ancestry from an admixture event at time ***T*** is maintained until the present. This low probability (never greater than 10$^{-10}$), argues against a model of *M. nasutus* ancestry in CACG or in DPRG being derived from a small number of introgression events.

    We explored numerous post-hoc strategies to ensure the robustness of our inference to idiosyncrasies in the identification of contiguous admixture

blocks. We first attempted to 'heal' physically close blocks into longer blocks. Specifically we connect admixture blocks separated by 0, 20,50 or 100 kb into one longer block (noted in column, 'Heal'). We also control for the potential of accidentally labeling regions of low divergence as introgressed by removing blocks whose length matches those identified in our allopatric samples, SLP or AHQT (noted in the column, 'Control for short blocks').

*Table S6)* Inference of introgression from *M. guttatus* to *M. nasutus* is robust to alternative thresholds of identifying outlier regions. The number of genomic regions where the outlier *M. nasutus* sample (or collection of samples) is closer to the *M. guttatus* sample of interest than is the average *M. nasutus* sample is presented before the slash and the total number of outlier regions with informative data is given after the slash. Note that the total number of outlier regions for a given *M. nasutus* sample may differ in comparisons to AHQT and SLP due to different patterns of the missingness of data. $\pi_S$ and $L$ denote the threshold $\pi_S$ and the sliding window size used to identity 20 kb outlier regions, respectively (see supplementary Analyses and Results for more details). Light grey shading represents the intermediate parameter values for identifying outliers regions reported in the main text.

*Table S7)* The negative relationship between recombination rate and divergence between *M. nasutus* and sympatric *M. guttatus* is not driven by sequencing depth or divergence to *M. dentilobus*. The first two columns present the percentage of fourfold degenerate sites that differ in the comparison (noted in the row heading) in regions with lower (low rec) and higher (high rec) than median recombination rates. The subsequent columns describe Spearman's $\rho$ and the P-value associated with this nonparametric correlation coefficient with alternative controls described in the supplement.

**Supplementary figure legends**

*Figure S1)* PSMC estimates of population diversity and divergence through time, showing the full range of recent potential maximum population sizes. Samples are identical to those in Figure 1E.

*Figure S2)* PSMC estimate of the split date between *M. nasutus* and southern *M. guttatus*. We infer speciation to occur when the between species curve (SLP x KOOT, gray) diverges from the southern *M. guttatus* curve (SLP x DPRG, blue). The black/dray gray line showing greater effective population size between *M. nasutus* and northern *M. guttatus* is shown for comparison.

*Figure S3)* PSMC inference shows *M. nasutus'* decline in effective population size. Effective population size through time is shown for pseudo-diploid genomes for all six pair-wise combinations of the four focal *M. nasutus* individuals. One intraspecific *M. guttatus* pair (AHQT x SLP, blue line) is shown for comparison.

*Figure S4)* PSMC suggests shared ancestry (in the form of a decrease in Ne) between *M. nasutus* and sympatric *M. guttatus* due to gene flow in sympatry. Population size through time is shown for pseudo-diploid genomes for pair-wise combinations of *M. guttatus* and/or *M. nasutus* individuals. Blue and violet = intraspecific *M. guttatus*. Black/gray = between species comparisons with allopatric *M. guttatus*. Brown/salmon and dark gold = interspecific comparisons with sympatric *M. guttatus*. Red = intraspecific *M. nasutus*.

*Figure S5)* PSMC suggests shared ancestry between *M. nasutus* and sympatric *M. guttatus*. This figure is zoomed out for scale. Samples are identical to those in Figure S4.

*Figure S6)* The allele frequency spectrum. The proportion of derived polymorphisms observed in one two or three (x-axis) *M. guttatus* (blue) and *M. nasutus* (red) samples. Filled and hatched bars describe four and zero-fold degenerate positions, respectively, while error bars indicate the upper and lower 2.5% of tails of the block bootstrap distribution.

*Figure S7)* Recent coalescence between focal samples and alternative *M. nasutus* across all chromosomes. Moving along each chromosome, we color genomic regions in which the focal individual and a *M. nasutus* sample (indicated by color) recently coalesce ($\pi_s \leq 0.5\%$). White regions coalesce more distantly in the past ($\pi_s > 0.5\%$) and grey regions indicate insufficient density of called sites. Tick marks on the x-axis indicate 1 megabase. For *M. nasutus* samples, colored regions represent common ancestry since the species split. Regions of *M. guttatus* genomes recently coalescing with *M. nasutus* likely represent recent introgression. Figure S7A presents recent coalescence across chromosomes 1-5, figure S7B presents recent coalescence across chromosomes 6-10, and S7C presents recent coalescence across chromosomes 10-14. Note that since we insist

on a high density of synonymous sites to evaluate 'recent coalescence' only a subset of our data informs our questions of recent coalescence. However, our HMM makes use of information from all genomic regions, and conditions on the density of sites with genotype data. Therefore, we can label introgression regions in places here we do not evaluate recent coalescence – i.e. purple lines indicating introgression into *M. guttatus* samples in gray regions of Figure S7.

Figure S8) Alternative Treemix analyses. We present all Treemix analyses varying data subset and number of admixture arrows. Left to Right: **(A)** Focal samples + the reference, rooted by the outgroup (MDENT) **(B)** Focal samples rooted by the outgroup (MDENT), or **(C)** All samples rooted by the outgroup (MDENT). Up to down **(1)** one, **(2)** two, **(3)** three, or **(4)** four admixture events.

Figure S9) The admixture block length distribution under alternative post-hoc 'healing' rules. The number of admixed blocks (as inferred by a greater than 95% posterior probability of *M. nasutus* ancestry from our HMM) longer than x. We joined two admixture blocks within **(A)** 0, **(B)** 20, **(C)** 50, or **(D)** 100 kb.

*Table S1)* Detailed information about the biology, geography, inbreeding and sequencing of each sample analyzed in this manuscript.

| ID | *Sp.* | State | (Long , Lat) | Ecology, etc. | Patry | LH [+] | # Gen Inbred | Read Length [♫] | Million Paired-End Reads | Machine [€] | Seq. Facility [¥] | NCBI SRA Accession No. |
|---|---|---|---|---|---|---|---|---|---|---|---|---|
| DENT-13007 | *dent* | AZ | unknown[#] | Inland | ? | PE | natural | 76 | 22.1 | GA-II | DoE JGI | SRX030541 |
| AHQT1.2 | *gutt* | WY | (-110.813, 44.431) | Inland, thermal | allo | AN | 1 | 100 | 55.8 | Hi-Seq | DoE JGI | SRX142379 |
| BOG10 | *gutt* | NV | (-118.805, 41.923) | Inland, hot springs | allo | PE | 3 | 76 | 28.0 | GA-II | DoE JGI | SRX030570 |
| CACG6 | *gutt* | WA | (-121.366, 45.710) | Inland | sym | AN | 3 | 100 | 85.2 | Hi-Seq | IGSP | NA |
| DPRG* | *gutt* | CA | (-120.344, 37.828) | Inland | sym | AN | 3 | 100 | 64.2 | Hi-Seq | DoE JGI | To be determined |
| DUN | *gutt* | OR | (-124.137, 43.893) | Coastal, dunes | allo | PE | >6 | 36 | 26.2 | GA-II | DoE JGI | SRX030973 SRX030974 |
| IM62 | *gutt* | OR | (-122.108, 44.481) | Inland | allo | AN | >10 | 100 | 103.2 | Hi-Seq | DoE JGI | SRX115898 |
| LMC24 | *gutt* | CA | (-123.084, 38.864) | Inland | allo | AN | 4 | 76 | 24.8 | GA-II | DoE JGI | SRX030680 |
| MAR3 | *gutt* | OR | (-123.294, 43.479) | Inland | allo | AN | 3 | 76 | 30.6 | GA-II | DoE JGI | SRX030542 |
| PED5 | *gutt* | AZ | (-110.130, 31.587) | Inland | allo | ? | 2 | 76 | 25.6 | GA-II | DoE JGI | SRX030544 |
| REM8-10 | *gutt* | CA | (-122.411, 38.860) | Inland, serpentine | allo | AN | 4 | 76 | 27.0 | GA-II | DoE JGI | SRX030546 |
| SLP19 | *gutt* | CA | (-120.462, 37.848) | Inland, serpentine | para | AN | ≥3 | 100 | 61.1 | Hi-Seq | DoE JGI | SRX142377 |
| SWB-S3-1-8 | *gutt* | CA | (-123.690, 39.036) | Coastal | allo | PE | 10 | 76 | 28.6 | GA-II | DoE JGI | SRX030679 |
| YJS6 | *gutt* | ID | (-114.585, 44.951) | Inland | allo | PE | 3 | 76 | 25.2 | GA-II | DoE JGI | SRX030545 |

| CACN9 | *nas* | WA | (-121.367, 45.711) | Inland, seep | sym | AN | natural | 100 | 58.4 | Hi-Seq | IGSP | NA |
| DPRN* | *nas* | CA | (-120.344, 37.829) | Inland | sym | AN | natural | 100 | 53.2 | Hi-Seq | IGSP | NA |
| KOOT | *nas* | MT | (-115.983, 48.104) | Inland | allo | AN | natural | 100 | 57.2 | Hi-Seq | IGSP | NA |
| NHN | *nas* | VI$^\$$ | (-124.160, 49.273) | Coastal, meadow | sym | AN | natural | 100 | 55.2 | Hi-Seq | IGSP | NA |
| SF5 | *nas* | OR | (-121.022, 45.264) | Inland | allo | AN | natural | 76 | 24.2 | GA-II | DoE JGI | SRX116529 |

\* DPRG and DPRN are referred to as MED and MEN in previous publications, respectively.
+ LH = Life History: AN = annual PE = perennial
# Maricopa County, AZ, USA
$ Vancouver Island, Canada
♫ Read length in base-pairs of a single end of paired end reads.
€ Sequencing Technology - Hi-Seq = Illumina Hi-Seq 2000; GA-II = Illumina Genome Analyzer II
¥ Sequencing facility - DoE JGI = U.S. Department of Energy Joint Genome Institute; IGSP = Duke Institute for Genome Sciences and Policy

Table S2) Pairwise sequence comparisons between all samples.

| Sample 1 | Sample 2 | $\pi_S$ | $\pi_N/\pi_S$ | Type of comparison | # Fo |
|---|---|---|---|---|---|
| CAC9N | KootN | 0.0091 | 0.2058 | M. nasutus | 2 |
| CAC9N | DPRN | 0.009 | 0.1946 | M. nasutus | 2 |
| CAC9N | NHN26 | 0.0112 | 0.1928 | M. nasutus | 2 |
| KootN | DPRN | 0.0083 | 0.2031 | M. nasutus | 2 |
| KootN | NHN26 | 0.009 | 0.1989 | M. nasutus | 2 |
| DPRN | NHN26 | 0.0108 | 0.1898 | M. nasutus | 2 |
| AHQT1G | CAC9N | 0.0544 | 0.1563 | M. nasutus X M. guttatus (North) | 2 |
| AHQT1G | KootN | 0.0547 | 0.1578 | M. nasutus X M. guttatus (North) | 2 |
| AHQT1G | DPRN | 0.0539 | 0.155 | M. nasutus X M. guttatus (North) | 2 |
| AHQT1G | NHN26 | 0.0537 | 0.1548 | M. nasutus X M. guttatus (North) | 2 |
| CAC6G | CAC9N | 0.0448 | 0.1611 | M. nasutus X M. guttatus (North) | 2 |
| CAC6G | KootN | 0.0459 | 0.1634 | M. nasutus X M. guttatus (North) | 2 |
| CAC6G | DPRN | 0.0453 | 0.1602 | M. nasutus X M. guttatus (North) | 2 |
| CAC6G | NHN26 | 0.0456 | 0.1601 | M. nasutus X M. guttatus (North) | 2 |
| CAC9N | DPRG | 0.0478 | 0.1566 | M. nasutus X M. guttatus (South) | 2 |
| CAC9N | SLP9G | 0.0507 | 0.1575 | M. nasutus X M. guttatus (South) | 2 |
| KootN | DPRG | 0.0481 | 0.1576 | M. nasutus X M. guttatus (South) | 2 |
| KootN | SLP9G | 0.0508 | 0.1589 | M. nasutus X M. guttatus (South) | 2 |
| DPRG | DPRN | 0.0473 | 0.154 | M. nasutus X M. guttatus (South) | 2 |
| DPRG | NHN26 | 0.0475 | 0.1544 | M. nasutus X M. guttatus (South) | 2 |
| DPRN | SLP9G | 0.0501 | 0.1562 | M. nasutus X M. guttatus (South) | 2 |
| NHN26 | SLP9G | 0.0503 | 0.1561 | M. nasutus X M. guttatus (South) | 2 |
| AHQT1G | CAC6G | 0.0398 | 0.161 | M. guttatus (North) | 2 |
| AHQT1G | DPRG | 0.0522 | 0.1542 | M. guttatus (South X North) | 2 |
| AHQT1G | SLP9G | 0.0534 | 0.1569 | M. guttatus (South X North) | 2 |
| CAC6G | DPRG | 0.0514 | 0.1579 | M. guttatus (South X North) | 2 |
| CAC6G | SLP9G | 0.0532 | 0.1593 | M. guttatus (South X North) | 2 |
| DPRG | SLP9G | 0.0446 | 0.1537 | M. guttatus (South) | 2 |
| CAC9N | SF5N | 0.0063 | 0.216 | M. nasutus | 1 |
| KootN | SF5N | 0.0073 | 0.2205 | M. nasutus | 1 |
| DPRN | SF5N | 0.008 | 0.2037 | M. nasutus | 1 |
| NHN26 | SF5N | 0.01 | 0.1953 | M. nasutus | 1 |
| CAC9N | Mdent | 0.0647 | 0.1576 | M. nasutus X M. dentilobus | 1 |
| KootN | Mdent | 0.0649 | 0.1585 | M. nasutus X M. dentilobus | 1 |
| Mdent | DPRN | 0.0641 | 0.1576 | M. nasutus X M. dentilobus | 1 |
| Mdent | NHN26 | 0.0641 | 0.1574 | M. nasutus X M. dentilobus | 1 |
| AHQT1G | SF5N | 0.0481 | 0.1547 | M. nasutus X M. guttatus (North) | 1 |
| BOG10G | CAC9N | 0.0503 | 0.1608 | M. nasutus X M. guttatus (North) | 1 |
| BOG10G | KootN | 0.0505 | 0.1624 | M. nasutus X M. guttatus (North) | 1 |
| BOG10G | DPRN | 0.0496 | 0.1603 | M. nasutus X M. guttatus (North) | 1 |
| BOG10G | NHN26 | 0.0495 | 0.1601 | M. nasutus X M. guttatus (North) | 1 |
| CAC6G | SF5N | 0.0399 | 0.1612 | M. nasutus X M. guttatus (North) | 1 |
| CAC9N | IM62.JGI | 0.0528 | 0.1563 | M. nasutus X M. guttatus (North) | 1 |
| CAC9N | IM62.LF | 0.0541 | 0.1564 | M. nasutus X M. guttatus (North) | 1 |
| CAC9N | MAR3G | 0.0536 | 0.1548 | M. nasutus X M. guttatus (North) | 1 |
| CAC9N | REM8G | 0.0453 | 0.1597 | M. nasutus X M. guttatus (North) | 1 |
| CAC9N | TSG3G | 0.053 | 0.1638 | M. nasutus X M. guttatus (North) | 1 |
| CAC9N | YJS6G | 0.0503 | 0.1592 | M. nasutus X M. guttatus (North) | 1 |
| IM62.JGI | KootN | 0.0532 | 0.1588 | M. nasutus X M. guttatus (North) | 1 |
| IM62.JGI | DPRN | 0.0519 | 0.1544 | M. nasutus X M. guttatus (North) | 1 |
| IM62.JGI | NHN26 | 0.0518 | 0.1546 | M. nasutus X M. guttatus (North) | 1 |
| IM62.LF | KootN | 0.0544 | 0.1587 | M. nasutus X M. guttatus (North) | 1 |
| IM62.LF | DPRN | 0.0532 | 0.1546 | M. nasutus X M. guttatus (North) | 1 |
| IM62.LF | NHN26 | 0.0529 | 0.1545 | M. nasutus X M. guttatus (North) | 1 |
| KootN | MAR3G | 0.0539 | 0.1571 | M. nasutus X M. guttatus (North) | 1 |
| KootN | REM8G | 0.0453 | 0.1628 | M. nasutus X M. guttatus (North) | 1 |
| KootN | TSG3G | 0.0531 | 0.164 | M. nasutus X M. guttatus (North) | 1 |
| KootN | YJS6G | 0.0504 | 0.1614 | M. nasutus X M. guttatus (North) | 1 |
| MAR3G | DPRN | 0.053 | 0.1543 | M. nasutus X M. guttatus (North) | 1 |
| MAR3G | NHN26 | 0.0528 | 0.1542 | M. nasutus X M. guttatus (North) | 1 |
| DPRN | REM8G | 0.0448 | 0.1592 | M. nasutus X M. guttatus (North) | 1 |
| DPRN | TSG3G | 0.0523 | 0.1631 | M. nasutus X M. guttatus (North) | 1 |
| DPRN | YJS6G | 0.0497 | 0.1585 | M. nasutus X M. guttatus (North) | 1 |
| NHN26 | REM8G | 0.045 | 0.159 | M. nasutus X M. guttatus (North) | 1 |
| NHN26 | TSG3G | 0.0524 | 0.1608 | M. nasutus X M. guttatus (North) | 1 |
| NHN26 | YJS6G | 0.0495 | 0.1591 | M. nasutus X M. guttatus (North) | 1 |
| CAC9N | DUNG | 0.0516 | 0.1543 | M. nasutus X M. guttatus (South) | 1 |
| CAC9N | LMC24G | 0.0435 | 0.1619 | M. nasutus X M. guttatus (South) | 1 |
| CAC9N | PED5G | 0.0467 | 0.1604 | M. nasutus X M. guttatus (South) | 1 |
| CAC9N | SWBG | 0.0473 | 0.1594 | M. nasutus X M. guttatus (South) | 1 |
| DUNG | KootN | 0.0517 | 0.1559 | M. nasutus X M. guttatus (South) | 1 |
| DUNG | DPRN | 0.0509 | 0.1532 | M. nasutus X M. guttatus (South) | 1 |
| DUNG | NHN26 | 0.0508 | 0.1528 | M. nasutus X M. guttatus (South) | 1 |
| KootN | LMC24G | 0.0435 | 0.1642 | M. nasutus X M. guttatus (South) | 1 |
| KootN | PED5G | 0.0466 | 0.1629 | M. nasutus X M. guttatus (South) | 1 |
| KootN | SWBG | 0.0475 | 0.1612 | M. nasutus X M. guttatus (South) | 1 |
| LMC24G | DPRN | 0.0429 | 0.1627 | M. nasutus X M. guttatus (South) | 1 |
| LMC24G | NHN26 | 0.0428 | 0.1626 | M. nasutus X M. guttatus (South) | 1 |
| DPRG | SF5N | 0.0432 | 0.154 | M. nasutus X M. guttatus (South) | 1 |
| DPRN | PED5G | 0.0461 | 0.1591 | M. nasutus X M. guttatus (South) | 1 |
| DPRN | SWBG | 0.0467 | 0.1591 | M. nasutus X M. guttatus (South) | 1 |
| NHN26 | PED5G | 0.046 | 0.1602 | M. nasutus X M. guttatus (South) | 1 |
| NHN26 | SWBG | 0.0466 | 0.1583 | M. nasutus X M. guttatus (South) | 1 |
| SF5N | SLP9G | 0.0459 | 0.1564 | M. nasutus X M. guttatus (South) | 1 |
| AHQT1G | Mdent | 0.06 | 0.1613 | M. guttatus (North) X M. dentilobus | 1 |
| CAC6G | Mdent | 0.0608 | 0.1621 | M. guttatus (North) X M. dentilobus | 1 |
| Mdent | DPRG | 0.0646 | 0.1583 | M. guttatus (South) X M. dentilobus | 1 |
| Mdent | SLP9G | 0.0647 | 0.1591 | M. guttatus (South) X M. dentilobus | 1 |
| AHQT1G | BOG10G | 0.0357 | 0.1599 | M. guttatus (North) | 1 |
| AHQT1G | IM62.JGI | 0.038 | 0.1568 | M. guttatus (North) | 1 |
| AHQT1G | IM62.LF | 0.0391 | 0.1573 | M. guttatus (North) | 1 |
| AHQT1G | MAR3G | 0.0349 | 0.1561 | M. guttatus (North) | 1 |
| AHQT1G | REM8G | 0.0456 | 0.1531 | M. guttatus (North) | 1 |
| AHQT1G | TSG3G | 0.0375 | 0.1639 | M. guttatus (North) | 1 |
| AHQT1G | YJS6G | 0.0349 | 0.1591 | M. guttatus (North) | 1 |
| BOG10G | CAC6G | 0.0397 | 0.1624 | M. guttatus (North) | 1 |
| CAC6G | IM62.JGI | 0.0413 | 0.1609 | M. guttatus (North) | 1 |
| CAC6G | IM62.LF | 0.0419 | 0.1628 | M. guttatus (North) | 1 |
| CAC6G | MAR3G | 0.0384 | 0.1597 | M. guttatus (North) | 1 |
| CAC6G | REM8G | 0.0453 | 0.1578 | M. guttatus (North) | 1 |
| CAC6G | TSG3G | 0.0411 | 0.1669 | M. guttatus (North) | 1 |
| CAC6G | YJS6G | 0.0394 | 0.1625 | M. guttatus (North) | 1 |
| AHQT1G | DUNG | 0.0444 | 0.1529 | M. guttatus (South X North) | 1 |
| AHQT1G | LMC24G | 0.0434 | 0.158 | M. guttatus (South X North) | 1 |
| AHQT1G | PED5G | 0.0417 | 0.1582 | M. guttatus (South X North) | 1 |
| AHQT1G | SWBG | 0.0426 | 0.1573 | M. guttatus (South X North) | 1 |
| BOG10G | DPRG | 0.0477 | 0.1582 | M. guttatus (South X North) | 1 |
| BOG10G | SLP9G | 0.049 | 0.1605 | M. guttatus (South X North) | 1 |
| CAC6G | DUNG | 0.0468 | 0.156 | M. guttatus (South X North) | 1 |
| CAC6G | LMC24G | 0.0432 | 0.1611 | M. guttatus (South X North) | 1 |
| CAC6G | PED5G | 0.0435 | 0.1624 | M. guttatus (South X North) | 1 |
| CAC6G | SWBG | 0.0443 | 0.1594 | M. guttatus (South X North) | 1 |
| IM62.JGI | DPRG | 0.0515 | 0.1545 | M. guttatus (South X North) | 1 |
| IM62.JGI | SLP9G | 0.0525 | 0.1569 | M. guttatus (South X North) | 1 |
| IM62.LF | DPRG | 0.0527 | 0.1545 | M. guttatus (South X North) | 1 |
| IM62.LF | SLP9G | 0.0541 | 0.1573 | M. guttatus (South X North) | 1 |
| MAR3G | DPRG | 0.0516 | 0.1528 | M. guttatus (South X North) | 1 |
| MAR3G | SLP9G | 0.0528 | 0.1558 | M. guttatus (South X North) | 1 |
| DPRG | REM8G | 0.042 | 0.152 | M. guttatus (South X North) | 1 |
| DPRG | TSG3G | 0.0509 | 0.1591 | M. guttatus (South X North) | 1 |
| DPRG | YJS6G | 0.0477 | 0.1574 | M. guttatus (South X North) | 1 |
| REM8G | SLP9G | 0.0437 | 0.1549 | M. guttatus (South X North) | 1 |
| SLP9G | TSG3G | 0.0519 | 0.1627 | M. guttatus (South X North) | 1 |
| SLP9G | YJS6G | 0.0489 | 0.1595 | M. guttatus (South X North) | 1 |
| DUNG | DPRG | 0.0488 | 0.1512 | M. guttatus (South) | 1 |
| DUNG | SLP9G | 0.05 | 0.1548 | M. guttatus (South) | 1 |
| LMC24G | DPRG | 0.0409 | 0.1542 | M. guttatus (South) | 1 |
| LMC24G | SLP9G | 0.0425 | 0.1595 | M. guttatus (South) | 1 |
| DPRG | PED5G | 0.0435 | 0.1564 | M. guttatus (South) | 1 |
| DPRG | SWBG | 0.044 | 0.1564 | M. guttatus (South) | 1 |
| PED5G | SLP9G | 0.0453 | 0.1598 | M. guttatus (South) | 1 |
| SLP9G | SWBG | 0.0455 | 0.1599 | M. guttatus (South) | 1 |
| Mdent | SF5N | 0.0595 | 0.1604 | M. nasutus X M. dentilobus | 0 |
| BOG10G | SF5N | 0.0446 | 0.1599 | M. nasutus X M. guttatus (North) | 0 |
| IM62.JGI | SF5N | 0.044 | 0.1523 | M. nasutus X M. guttatus (North) | 0 |
| IM62.LF | SF5N | 0.045 | 0.1538 | M. nasutus X M. guttatus (North) | 0 |
| MAR3G | SF5N | 0.0471 | 0.1532 | M. nasutus X M. guttatus (North) | 0 |
| REM8G | SF5N | 0.0412 | 0.1607 | M. nasutus X M. guttatus (North) | 0 |
| SF5N | TSG3G | 0.0465 | 0.166 | M. nasutus X M. guttatus (North) | 0 |
| SF5N | YJS6G | 0.0446 | 0.1579 | M. nasutus X M. guttatus (North) | 0 |
| DUNG | SF5N | 0.0461 | 0.1526 | M. nasutus X M. guttatus (South) | 0 |
| LMC24G | SF5N | 0.0388 | 0.1637 | M. nasutus X M. guttatus (South) | 0 |
| PED5G | SF5N | 0.042 | 0.1608 | M. nasutus X M. guttatus (South) | 0 |
| SF5N | SWBG | 0.0424 | 0.1584 | M. nasutus X M. guttatus (South) | 0 |
| BOG10G | Mdent | 0.0558 | 0.1694 | M. guttatus (North) X M. dentilobus | 0 |
| IM62.JGI | Mdent | 0.054 | 0.1627 | M. guttatus (North) X M. dentilobus | 0 |
| IM62.LF | Mdent | 0.0557 | 0.1615 | M. guttatus (North) X M. dentilobus | 0 |
| MAR3G | Mdent | 0.0578 | 0.1618 | M. guttatus (North) X M. dentilobus | 0 |
| Mdent | REM8G | 0.0581 | 0.1642 | M. guttatus (North) X M. dentilobus | 0 |
| Mdent | TSG3G | 0.0591 | 0.164 | M. guttatus (North) X M. dentilobus | 0 |
| Mdent | YJS6G | 0.0567 | 0.1662 | M. guttatus (North) X M. dentilobus | 0 |
| DUNG | Mdent | 0.0599 | 0.1598 | M. guttatus (South) X M. dentilobus | 0 |
| LMC24G | Mdent | 0.0574 | 0.1652 | M. guttatus (South) X M. dentilobus | 0 |
| Mdent | PED5G | 0.0568 | 0.1637 | M. guttatus (South) X M. dentilobus | 0 |
| Mdent | SWBG | 0.0572 | 0.166 | M. guttatus (South) X M. dentilobus | 0 |
| BOG10G | IM62.JGI | 0.0345 | 0.1598 | M. guttatus (North) | 0 |
| BOG10G | IM62.LF | 0.0354 | 0.162 | M. guttatus (North) | 0 |
| BOG10G | MAR3G | 0.0345 | 0.161 | M. guttatus (North) | 0 |
| BOG10G | REM8G | 0.042 | 0.1631 | M. guttatus (North) | 0 |
| BOG10G | TSG3G | 0.0372 | 0.1789 | M. guttatus (North) | 0 |
| BOG10G | YJS6G | 0.0342 | 0.1662 | M. guttatus (North) | 0 |
| IM62.JGI | IM62.LF | 0.0003 | 0.2433 | M. guttatus (North) | 0 |
| IM62.JGI | MAR3G | 0.0319 | 0.1545 | M. guttatus (North) | 0 |
| IM62.JGI | REM8G | 0.0421 | 0.1518 | M. guttatus (North) | 0 |
| IM62.JGI | TSG3G | 0.0339 | 0.1603 | M. guttatus (North) | 0 |
| IM62.JGI | YJS6G | 0.0342 | 0.1564 | M. guttatus (North) | 0 |
| IM62.LF | MAR3G | 0.033 | 0.1553 | M. guttatus (North) | 0 |
| IM62.LF | REM8G | 0.0425 | 0.1551 | M. guttatus (North) | 0 |
| IM62.LF | TSG3G | 0.0352 | 0.1536 | M. guttatus (North) | 0 |
| IM62.LF | YJS6G | 0.0353 | 0.1554 | M. guttatus (North) | 0 |
| MAR3G | REM8G | 0.0443 | 0.154 | M. guttatus (North) | 0 |
| MAR3G | TSG3G | 0.035 | 0.1613 | M. guttatus (North) | 0 |
| MAR3G | YJS6G | 0.0355 | 0.1575 | M. guttatus (North) | 0 |
| REM8G | TSG3G | 0.042 | 0.1797 | M. guttatus (North) | 0 |
| REM8G | YJS6G | 0.0423 | 0.1595 | M. guttatus (North) | 0 |
| TSG3G | YJS6G | 0.0386 | 0.1695 | M. guttatus (North) | 0 |
| BOG10G | DUNG | 0.0418 | 0.1568 | M. guttatus (South X North) | 0 |
| BOG10G | LMC24G | 0.0394 | 0.1663 | M. guttatus (South X North) | 0 |
| BOG10G | PED5G | 0.0379 | 0.1675 | M. guttatus (South X North) | 0 |
| BOG10G | SWBG | 0.0395 | 0.162 | M. guttatus (South X North) | 0 |
| DUNG | IM62.JGI | 0.0433 | 0.1508 | M. guttatus (South X North) | 0 |
| DUNG | IM62.LF | 0.0441 | 0.1512 | M. guttatus (South X North) | 0 |
| DUNG | MAR3G | 0.0423 | 0.1525 | M. guttatus (South X North) | 0 |
| DUNG | REM8G | 0.0391 | 0.156 | M. guttatus (South X North) | 0 |
| DUNG | TSG3G | 0.038 | 0.158 | M. guttatus (South X North) | 0 |
| DUNG | YJS6G | 0.0414 | 0.154 | M. guttatus (South X North) | 0 |
| IM62.JGI | LMC24G | 0.0391 | 0.1569 | M. guttatus (South X North) | 0 |
| IM62.JGI | PED5G | 0.0392 | 0.1572 | M. guttatus (South X North) | 0 |
| IM62.JGI | SWBG | 0.0397 | 0.1548 | M. guttatus (South X North) | 0 |
| IM62.LF | LMC24G | 0.0401 | 0.1589 | M. guttatus (South X North) | 0 |
| IM62.LF | PED5G | 0.0403 | 0.156 | M. guttatus (South X North) | 0 |
| IM62.LF | SWBG | 0.0409 | 0.1544 | M. guttatus (South X North) | 0 |
| LMC24G | MAR3G | 0.042 | 0.1577 | M. guttatus (South X North) | 0 |
| LMC24G | REM8G | 0.03 | 0.1767 | M. guttatus (South X North) | 0 |
| LMC24G | TSG3G | 0.0415 | 0.1764 | M. guttatus (South X North) | 0 |
| LMC24G | YJS6G | 0.0392 | 0.1668 | M. guttatus (South X North) | 0 |
| MAR3G | PED5G | 0.0409 | 0.1578 | M. guttatus (South X North) | 0 |
| MAR3G | SWBG | 0.0416 | 0.1555 | M. guttatus (South X North) | 0 |
| PED5G | REM8G | 0.0397 | 0.1628 | M. guttatus (South X North) | 0 |
| PED5G | TSG3G | 0.041 | 0.1768 | M. guttatus (South X North) | 0 |
| PED5G | YJS6G | 0.0382 | 0.1641 | M. guttatus (South X North) | 0 |
| REM8G | SWBG | 0.0349 | 0.1671 | M. guttatus (South X North) | 0 |
| SWBG | TSG3G | 0.0366 | 0.1821 | M. guttatus (South X North) | 0 |
| SWBG | YJS6G | 0.0388 | 0.1614 | M. guttatus (South X North) | 0 |
| DUNG | LMC24G | 0.0372 | 0.1621 | M. guttatus (South) | 0 |
| DUNG | PED5G | 0.0409 | 0.1566 | M. guttatus (South) | 0 |
| DUNG | SWBG | 0.0277 | 0.1661 | M. guttatus (South) | 0 |
| LMC24G | PED5G | 0.0376 | 0.1667 | M. guttatus (South) | 0 |
| LMC24G | SWBG | 0.032 | 0.1732 | M. guttatus (South) | 0 |
| PED5G | SWBG | 0.0372 | 0.1649 | M. guttatus (South) | 0 |

Table S3) A summary of pairwise sequence diversity in comparisons between all samples.

| Comparison | # of focal samples | $\pi_S$ | $\pi_N/\pi_S$ |
|---|---|---|---|
| M. guttatus (North) | 0 | 0.032881114 | 0.158215329 |
| M. guttatus (North) | 1 | 0.03905176 | 0.159231447 |
| M. guttatus (North) | 2 | 0.03975362 | 0.160967574 |
| M. guttatus (North) X M. dentilobus | 0 | 0.056176252 | 0.163768857 |
| M. guttatus (North) X M. dentilobus | 1 | 0.060400169 | 0.161706935 |
| M. guttatus (South X North) | 0 | 0.040716802 | 0.156685768 |
| M. guttatus (South X North) | 1 | 0.048184677 | 0.156386183 |
| M. guttatus (South X North) | 2 | 0.052557792 | 0.157113957 |
| M. guttatus (South) | 0 | 0.034973709 | 0.163133394 |
| M. guttatus (South) | 1 | 0.046413063 | 0.155447195 |
| M. guttatus (South) | 2 | 0.044593728 | 0.153736849 |
| M. guttatus (South) X M. dentilobus | 0 | 0.058394551 | 0.162545546 |
| M. guttatus (South) X M. dentilobus | 1 | 0.064657437 | 0.158693178 |
| M. nasutus | 1 | 0.007883245 | 0.207432369 |
| M. nasutus | 2 | 0.009565595 | 0.197091862 |
| M. nasutus X M. dentilobus | 0 | 0.059518884 | 0.160402991 |
| M. nasutus X M. dentilobus | 1 | 0.064453912 | 0.157786429 |
| M. nasutus X M. guttatus (North) | 0 | 0.044767503 | 0.155487085 |
| M. nasutus X M. guttatus (North) | 1 | 0.050955018 | 0.157330052 |
| M. nasutus X M. guttatus (North) | 2 | 0.049721701 | 0.158399731 |
| M. nasutus X M. guttatus (South) | 0 | 0.043482976 | 0.156725938 |
| M. nasutus X M. guttatus (South) | 1 | 0.047854641 | 0.157210047 |
| M. nasutus X M. guttatus (South) | 2 | 0.049053125 | 0.156452924 |

*Table S4)* After removing regions of recent introgression, interspecific divergence corresponds to the topology of the neighbor-joining tree.

| Mean # of synonymous sequence diffs % | | | Including introgression regions | | | | | | | |
|---|---|---|---|---|---|---|---|---|---|---|
| | | | *M. guttatus* | | | | *M. nasutus* | | | |
| | | | AHQT | CACG | DPRG | SLP | NHN | DPRN | KootN | CACN |
| Excluding introgression | *M. guttatus* | AHQT | | 3.98 | 5.22 | 5.34 | 5.37 | 5.39 | 5.47 | 5.44 |
| | | CACG | 3.65 | | 5.14 | 5.32 | 4.56 | 4.53 | 4.59 | 4.48 |
| | | DPRG | 5.26 | 5.29 | | 4.46 | 4.75 | 4.73 | 4.81 | 4.78 |
| | | SLP | 5.37 | 5.42 | 4.48 | | 5.03 | 5.01 | 5.08 | 5.07 |
| | *M. nasutus* | NHN | 5.39 | 5.38 | 4.95 | 5.06 | | 1.08 | 0.90 | 1.12 |
| | | DPRN | 5.41 | 5.40 | 4.93 | 5.05 | 1.08 | | 0.83 | 0.90 |
| | | Koot | 5.49 | 5.48 | 5.02 | 5.12 | 0.90 | 0.83 | | 0.91 |
| | | CACN | 5.46 | 5.44 | 4.98 | 5.10 | 1.12 | 0.90 | 0.91 | |

*Table S5)* Summary of the admixture block-length distribution and the robustness of our inference of admixture history.

| Focal Sample | Control for short blocks | Heal | # of blocks | Length (kb) | Mean Length (cM) | Sdv length (cM) | *T* - Time since admixture | P(exp) | Prob (block inherited from point event) |
|---|---|---|---|---|---|---|---|---|---|
| CACG | none | 0 | 227 | 29977 | 0.74 | 1.04 | 135 | <0.001 | 9.40E-38 |
| | | 20 | 180 | 30554 | 0.95 | 1.45 | 105 | <0.001 | 7.68E-29 |
| | | 50 | 138 | 32031 | 1.3 | 2.5 | 77 | <0.001 | 1.75E-20 |
| | | 100 | 112 | 33873 | 1.7 | 3.19 | 59 | <0.001 | 3.20E-15 |
| | ahqt | 0 | 204 | 29352 | 0.81 | 1.08 | 124 | <0.001 | 1.88E-34 |
| | | 20 | 157 | 29928 | 1.07 | 1.51 | 93 | <0.001 | 1.98E-25 |
| | | 50 | 115 | 31412 | 1.53 | 2.68 | 65 | <0.001 | 4.43E-17 |
| | | 100 | 90 | 33182 | 2.07 | 3.45 | 48 | <0.001 | 4.06E-12 |
| | Slp | 0 | 171 | 29021 | 0.95 | 1.12 | 105 | 0.03 | 7.58E-29 |
| | | 20 | 130 | 29450 | 1.27 | 1.59 | 79 | 0.04 | 4.85E-21 |
| | | 50 | 90 | 30770 | 1.92 | 2.91 | 52 | 0.04 | 3.16E-13 |
| | | 100 | 67 | 32235 | 2.7 | 3.81 | 37 | 0.06 | 7.88E-09 |
| DPRG | None | 0 | 350 | 6484 | 0.1 | 0.13 | 962 | <0.001 | 6.89E-286 |
| | | 20 | 306 | 6979 | 0.13 | 0.19 | 781 | <0.001 | 1.31E-231 |
| | | 50 | 269 | 8194 | 0.17 | 0.25 | 585 | <0.001 | 1.26E-172 |
| | | 100 | 231 | 11101 | 0.27 | 0.39 | 371 | <0.001 | 2.47E-108 |
| | Ahqt | 0 | 327 | 5848 | 0.1 | 0.13 | 997 | <0.001 | 2.87E-296 |
| | | 20 | 283 | 6348 | 0.13 | 0.18 | 795 | <0.001 | 1.51E-235 |
| | | 50 | 246 | 7560 | 0.17 | 0.25 | 580 | <0.001 | 4.44E-171 |
| | | 100 | 209 | 10412 | 0.28 | 0.4 | 358 | <0.001 | 2.12E-104 |
| | Slp | 0 | 294 | 5622 | 0.11 | 0.14 | 932 | <0.001 | 7.28E-277 |
| | | 20 | 256 | 6041 | 0.13 | 0.19 | 755 | 0.02 | 9.58E-224 |
| | | 50 | 221 | 7212 | 0.18 | 0.26 | 546 | <0.001 | 6.25E-161 |
| | | 100 | 186 | 9855 | 0.3 | 0.4 | 336 | <0.001 | 5.41E-98 |
| SLP | none | 0 | 56 | 867 | 0.09 | 0.11 | 1151 | 0.1 | 0 |
| | | 20 | 50 | 932 | 0.1 | 0.18 | 956 | 0.11 | 4.09E-284 |
| | | 50 | 48 | 982 | 0.11 | 0.18 | 871 | 0.11 | 1.43E-258 |
| | | 100 | 45 | 1192 | 0.15 | 0.3 | 673 | 0.06 | 5.60E-199 |
| | ahqt | 0 | 33 | 330 | 0.06 | 0.09 | 1782 | 0.37 | 0 |
| | | 20 | 27 | 331 | 0.07 | 0.14 | 1454 | 0.34 | 0 |
| | | 50 | 25 | 376 | 0.08 | 0.14 | 1185 | 0.3 | 0 |
| | | 100 | 23 | 559 | 0.14 | 0.36 | 733 | 0.17 | 3.80E-217 |
| AHQT | none | 0 | 23 | 630 | 0.15 | 0.21 | 651 | 0.37 | 2.55E-192 |
| | | 20 | 23 | 630 | 0.15 | 0.21 | 651 | 0.37 | 2.55E-192 |
| | | 50 | 23 | 630 | 0.15 | 0.21 | 651 | 0.34 | 2.55E-192 |
| | | 100 | 22 | 689 | 0.18 | 0.26 | 569 | 0.12 | 8.10E-168 |

*Table S6)* Inference of introgression from *M. guttatus* to *M. nasutus* is robust to alternative thresholds of identifying outlier regions.

| A | *L* | $\pi_S\%$ | DPRN | NHN | Koot | CACN | Northern_nas |
|---|---|---|---|---|---|---|---|
| To AHQ | 5 | 0.5 | 47/102 (0.814) | **138/231 (0.002)** | 31/60 (0.449) | 101/207 (0.662) | **270/498 (0.033)** |
| | 5 | 1.0 | 40/80 (0.544) | **120/197 (0.001)** | 25/53 (0.708) | 81/154 (0.286) | **226/404 (0.01)** |
| | 5 | 2.0 | 28/56 (0.317) | **90/133 (0.02)** | 17/38 (0.76) | 59/106 (0.636) | 166/277 (0.144) |
| | 10 | 0.5 | 71/154 (0.853) | **164/290 (0.015)** | 39/80 (0.631) | 123/243 (0.449) | 326/613 (0.062) |
| | 10 | 1.0 | 68/134 (0.466) | **141/249 (0.021)** | 36/67 (0.313) | 95/174 (0.128) | **272/490 (0.008)** |
| | 10 | 2.0 | 43/89 (0.664) | 90/165 (0.138) | 32/54 (0.11) | 61/116 (0.321) | **183/335 (0.05)** |
| | 20 | 0.5 | 126/247 (0.4) | **245/453 (0.045)** | 40/105 (0.995) | 186/348 (0.109) | 471/906 (0.122) |
| | 20 | 1.0 | 112/216 (0.317) | **222/402 (0.02)** | 46/98 (0.76) | 146/297 (0.636) | 414/797 (0.144) |
| | 20 | 2.0 | 78/139 (0.087) | 147/273 (0.113) | 42/77 (0.247) | 107/198 (0.143) | **296/548 (0.033)** |
| B | *L* | $\pi_S$ | DPRN | NHN | Koot | CACN | Northern_nas |
| To SLP | 5 | 0.5 | 36/71 (0.5) | 85/172 (0.59) | 13/32 (0.892) | 50/111 (0.873) | 148/315 (0.87) |
| | 5 | 1.0 | 27/56 (0.656) | 72/136 (0.274) | 17/34 (0.568) | 37/79 (0.75) | 126/249 (0.45) |
| | 5 | 2.0 | 16/39 (0.432) | 49/91 (0.596) | 13/24 (0.551) | 26/52 (0.95) | 88/167 (0.872) |
| | 10 | 0.5 | 48/99 (0.656) | 112/217 (0.342) | 18/47 (0.96) | 72/153 (0.79) | 202/417 (0.754) |
| | 10 | 1.0 | 42/82 (0.456) | 97/186 (0.304) | 19/46 (0.908) | 51/108 (0.75) | 167/340 (0.648) |
| | 10 | 2.0 | 22/52 (0.894) | 62/124 (0.536) | 17/33 (0.5) | 34/68 (0.548) | 113/225 (0.5) |
| | 20 | 0.5 | 73/159 (0.867) | 156/305 (0.366) | 23/61 (0.98) | 111/224 (0.579) | 290/590 (0.675) |
| | 20 | 1.0 | 70/137 (0.432) | 133/269 (0.596) | 30/60 (0.551) | 87/196 (0.95) | 250/525 (0.872) |
| | 20 | 2.0 | 45/91 (0.583) | 90/180 (0.53) | 24/48 (0.557) | 44/111 (0.989) | 158/339 (0.904) |

*Table S7)* The negative relationship between recombination rate and divergence between *M. nasutus* and sympatric *M. guttatus* is not driven by sequencing depth or divergence to *M. dentilobus*.

| Differentiation by recombination rate | Mean # of pairwise sequence differences (%) | | No control | | Control for mutation (divergence to M. dent) | | Control for depth (at synonymous sites) | | Control for mutation and depth | |
|---|---|---|---|---|---|---|---|---|---|---|
| | low rec | high rec | P-value | ρ | P-value | ρ | P-value | ρ | P-value | ρ |
| Within *M. nasutus* | 0.70 | 0.81 | 0.1228 | 0.0370 | 0.2553 | 0.0273 | 0.1439 | 0.0350 | 0.2375 | 0.0283 |
| Within northern *M. guttatus* | 4.03 | 4.08 | 0.6744 | -0.0101 | 0.6764 | -0.0100 | 0.5734 | -0.0135 | 0.5865 | -0.0130 |
| Within southern *M. guttatus* | 4.59 | 4.62 | 0.5353 | -0.0149 | 0.8785 | 0.0037 | 0.4509 | -0.0181 | 0.9571 | -0.0013 |
| Within all *M. guttatus* | 5.17 | 5.09 | 0.2441 | -0.0279 | 0.6572 | -0.0106 | 0.2044 | -0.0304 | 0.5191 | -0.0155 |
| Between CACG and *M. nasutus* | 5.48 | 5.14 | 0.0027 | -0.0718 | 0.0070 | -0.0646 | 0.0022 | -0.0732 | 0.0064 | -0.0652 |
| Between DPRN and *M. nasutus* | 5.11 | 4.93 | 0.0008 | -0.0800 | 0.0022 | -0.0732 | 0.0006 | -0.0822 | 0.0016 | -0.0754 |
| Between SLP and *M. nasutus* | 5.28 | 5.15 | 0.0297 | -0.0521 | 0.0458 | -0.0479 | 0.0241 | -0.0540 | 0.0354 | -0.0504 |
| Between AHQT and *M. nasutus* | 5.63 | 5.54 | 0.2768 | -0.0261 | 0.5338 | -0.0149 | 0.2388 | -0.0282 | 0.4551 | -0.0179 |

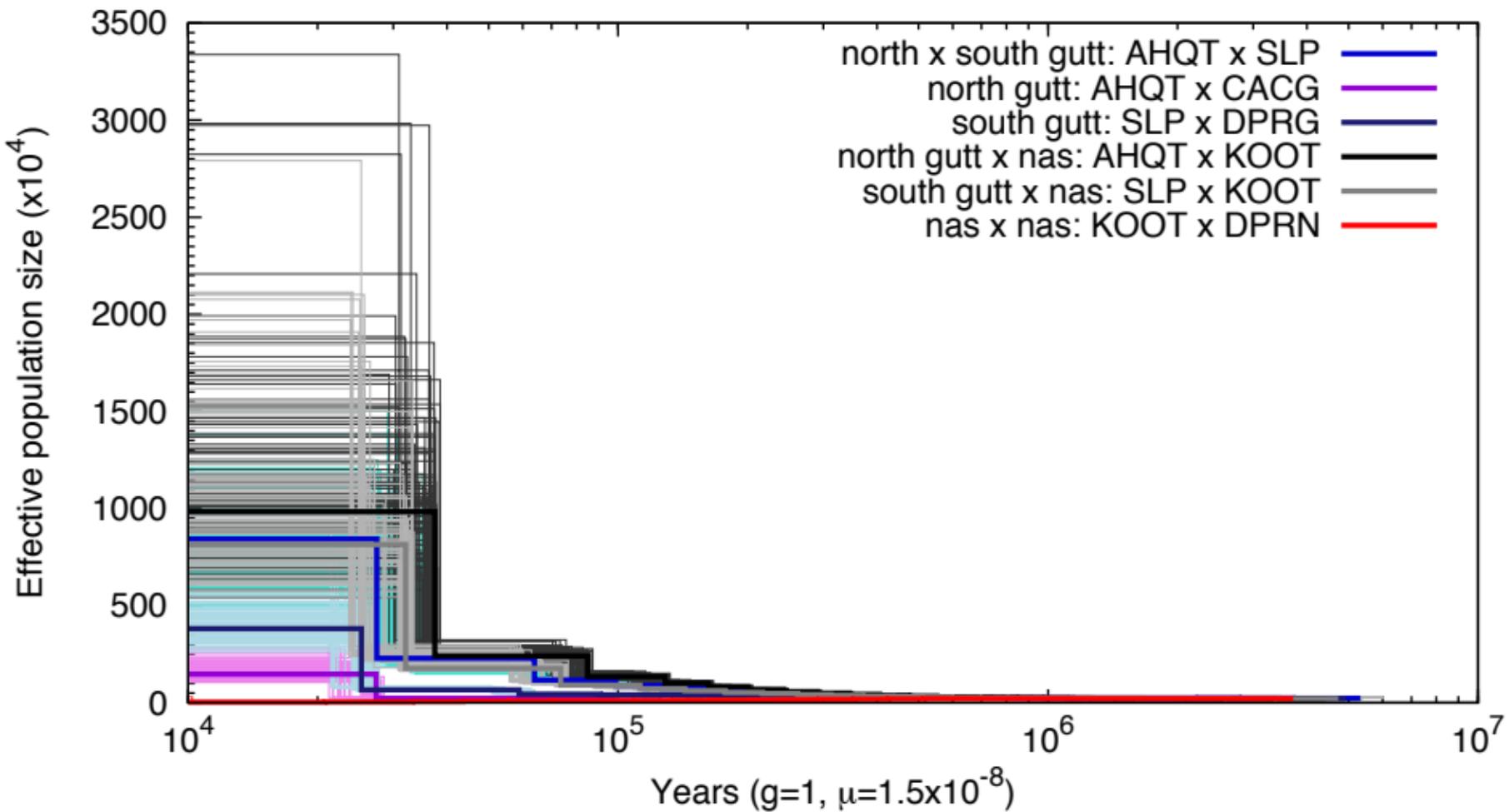

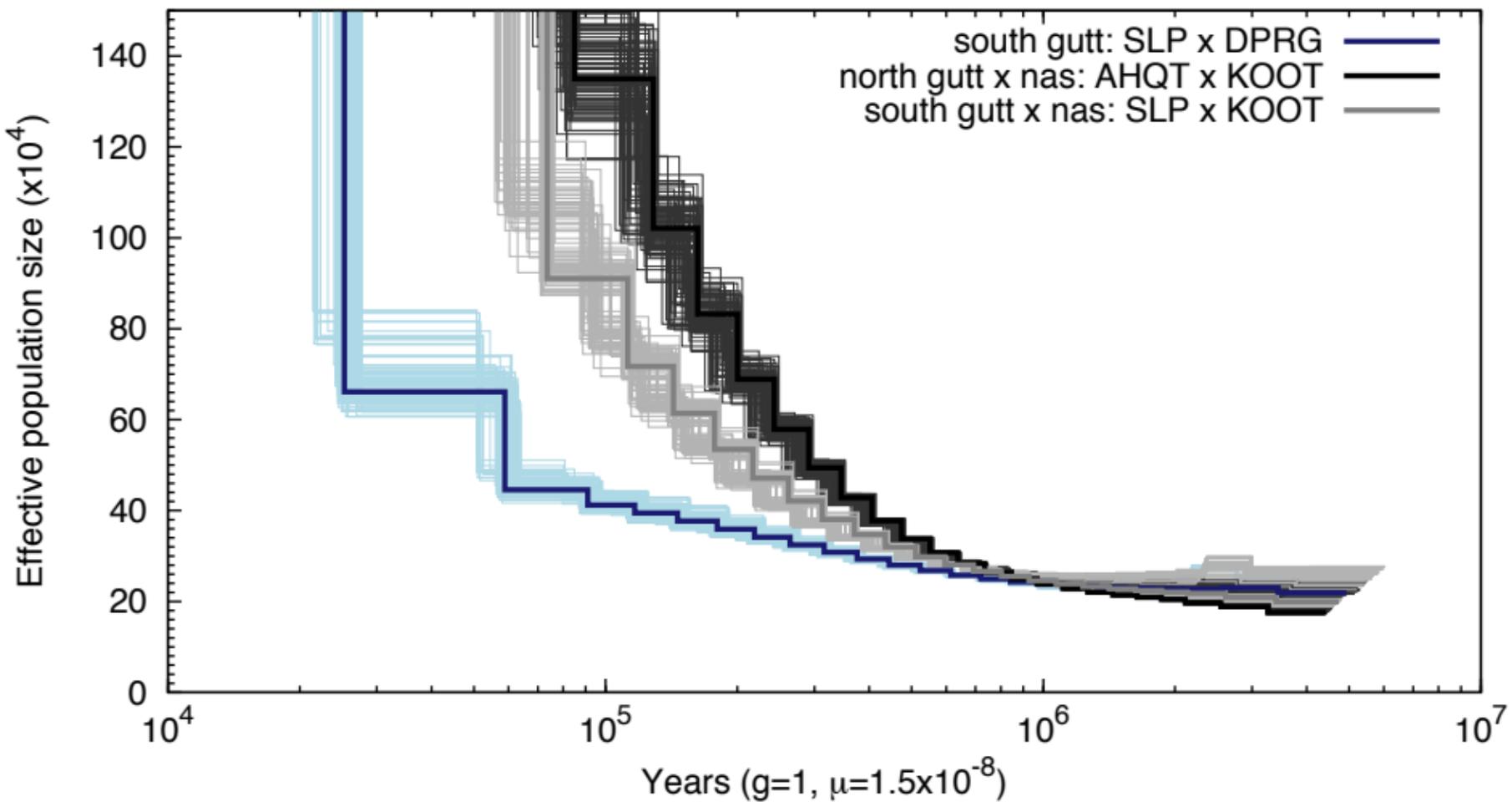

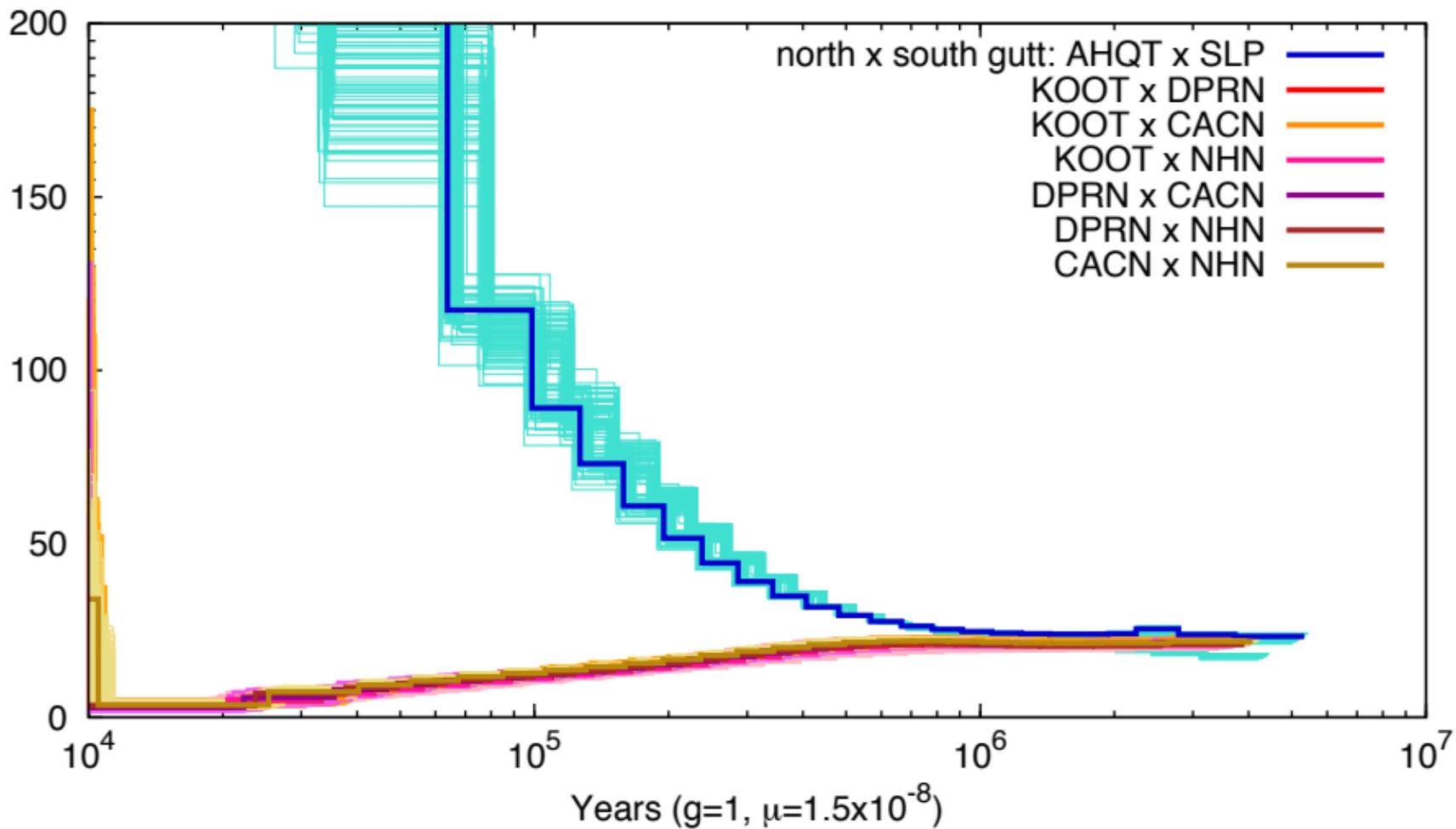

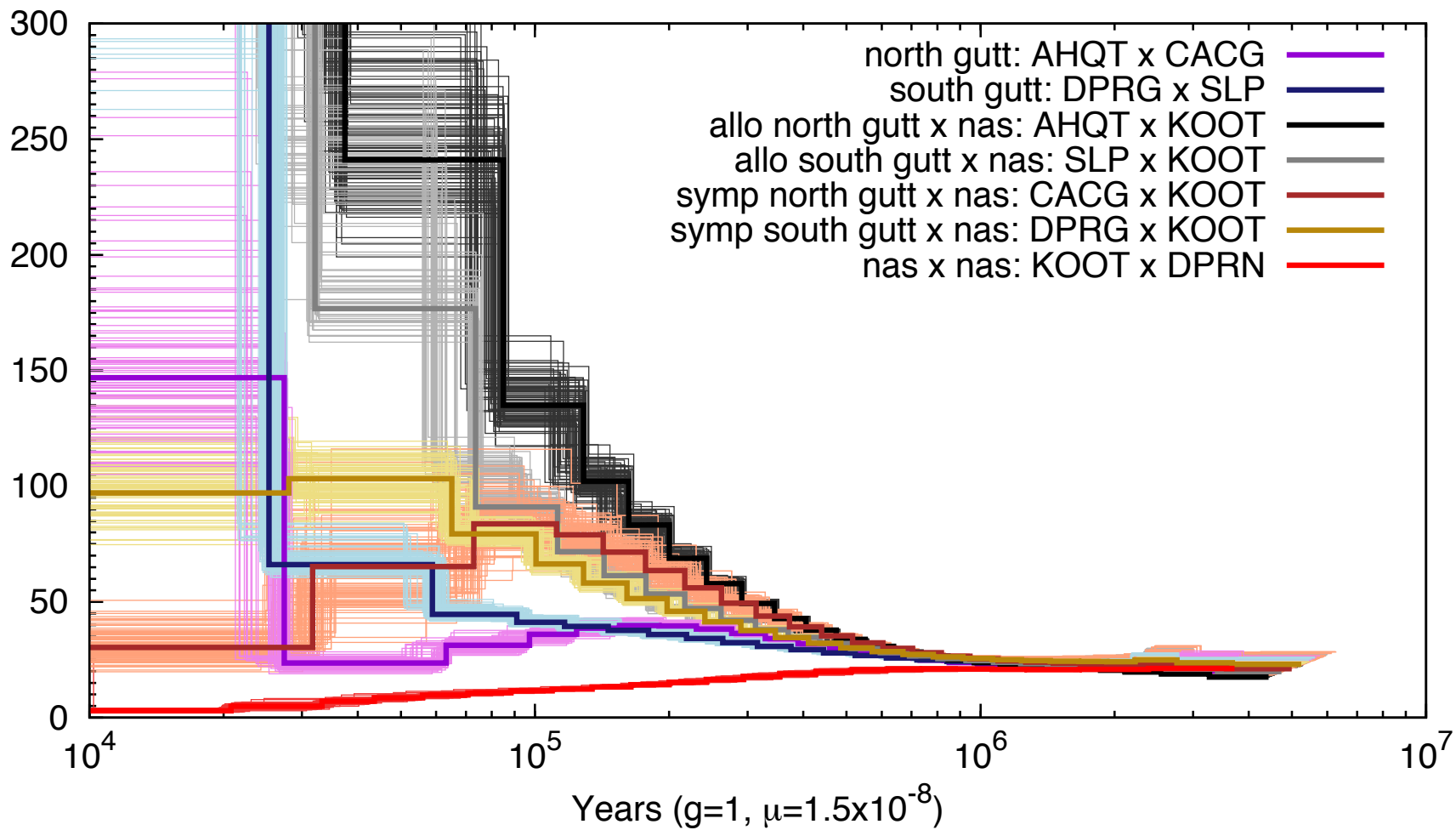

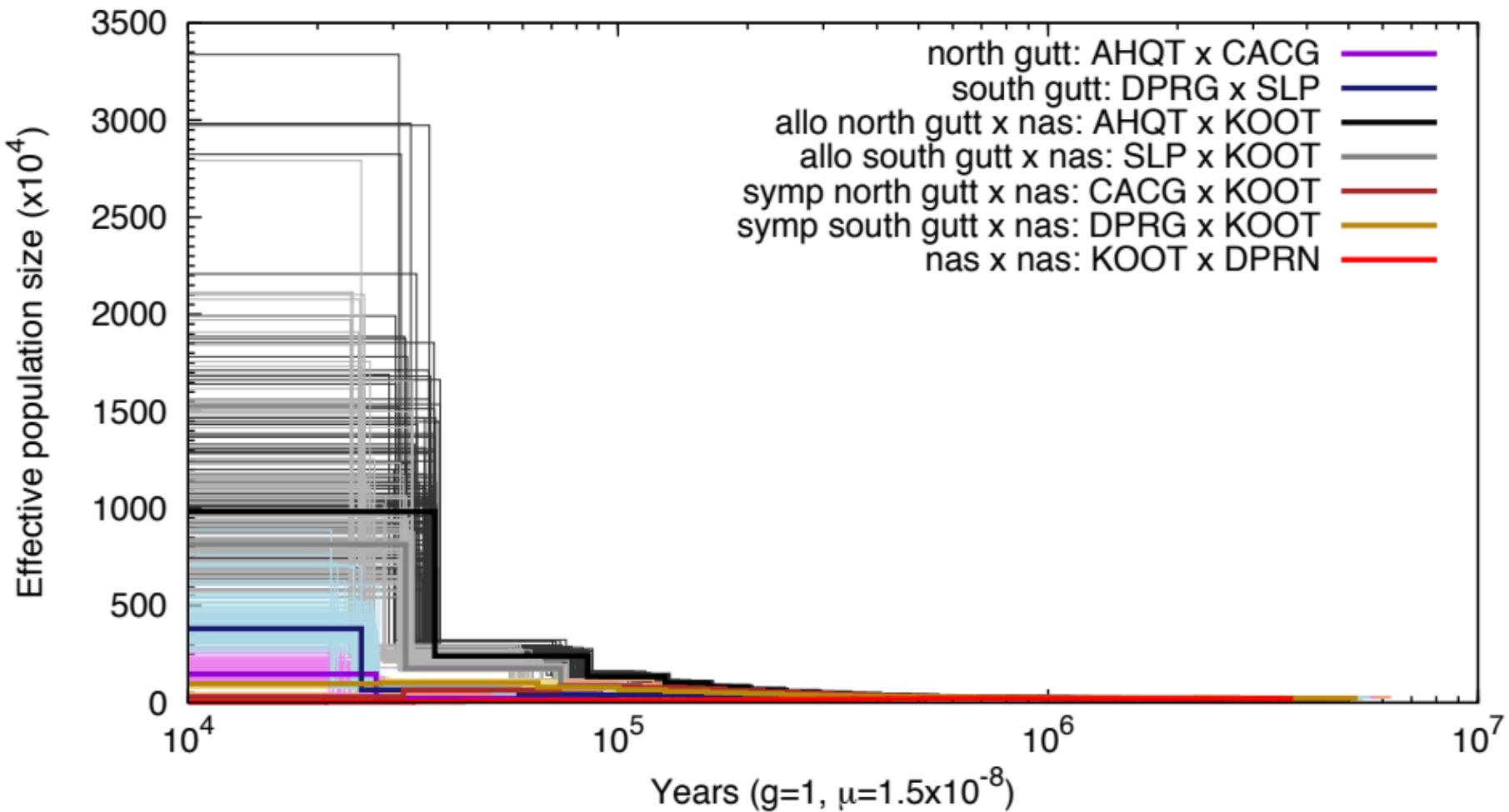

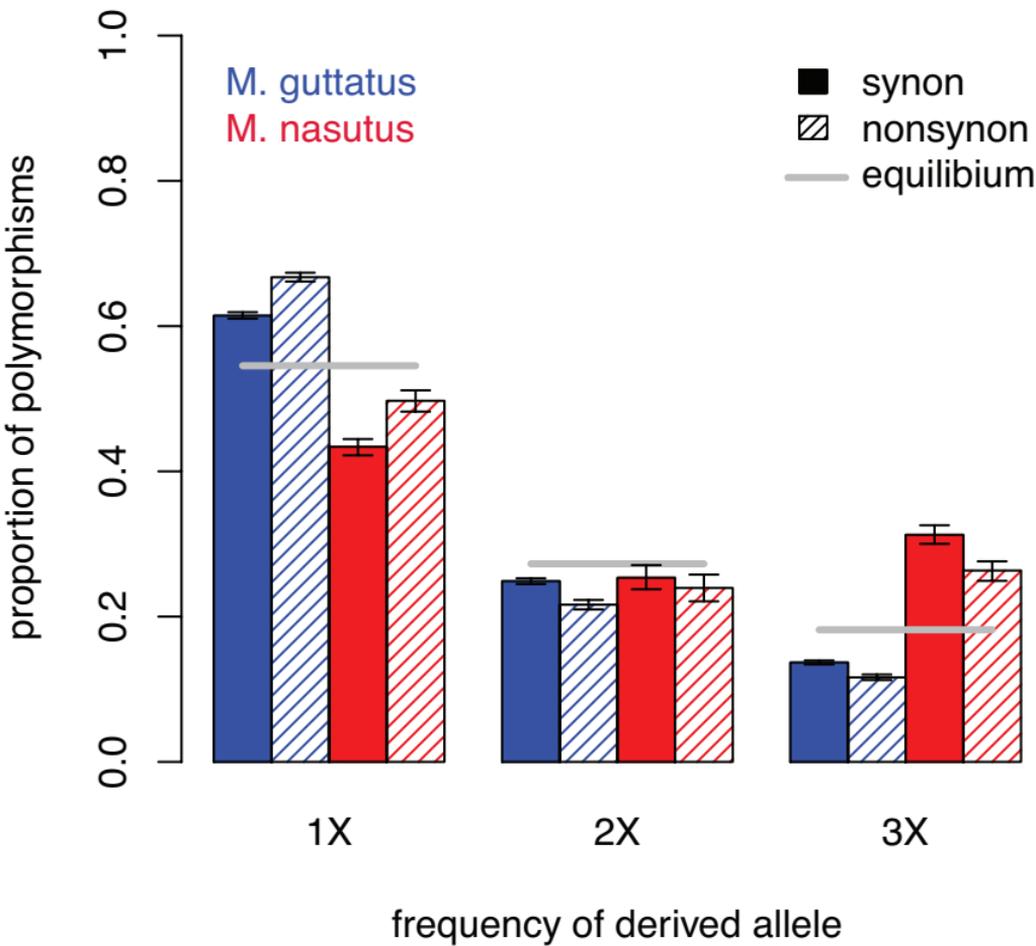

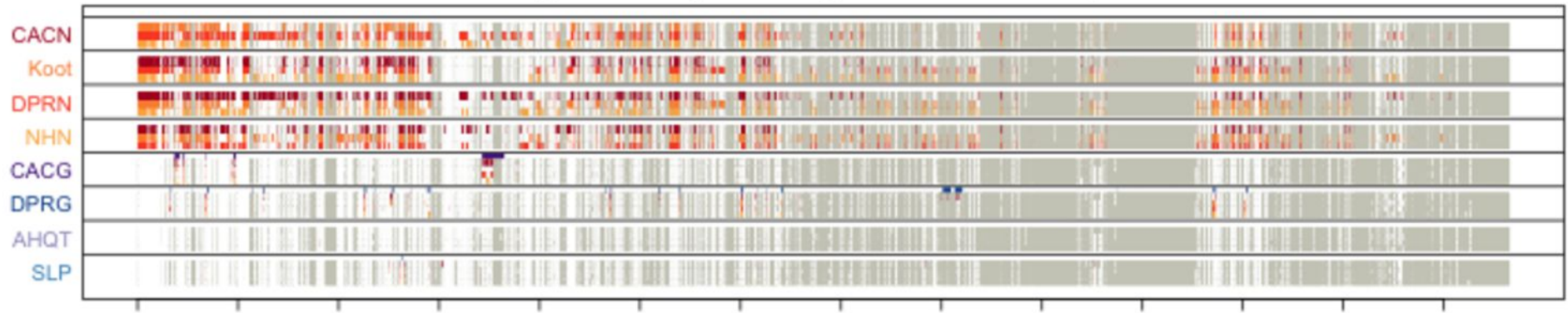
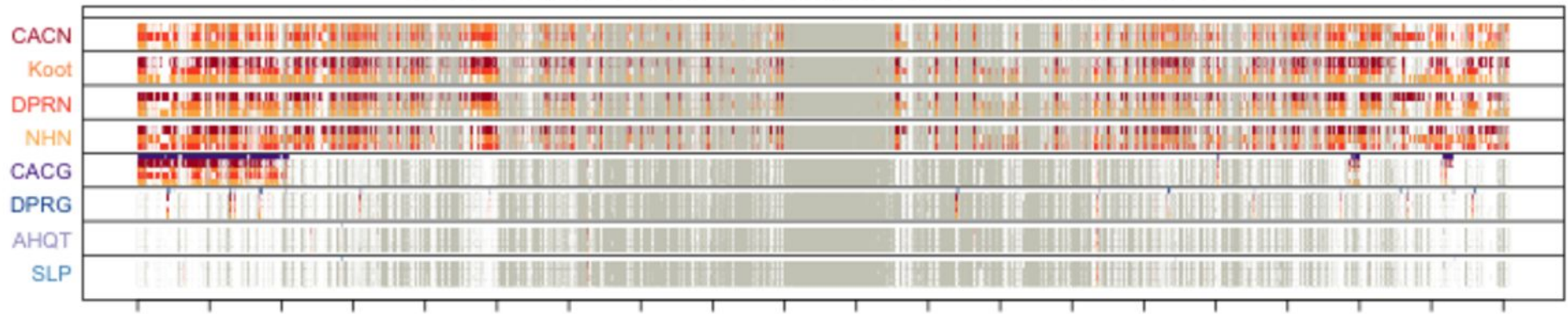
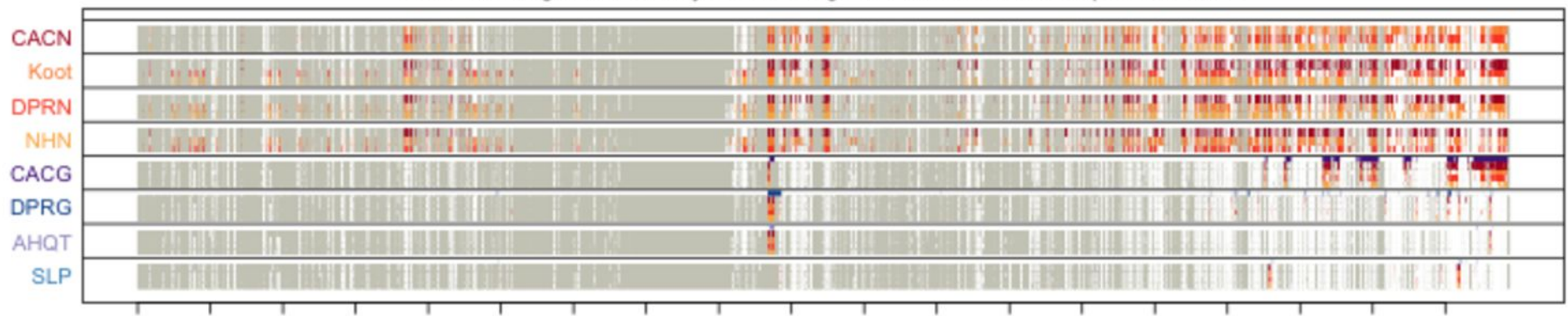
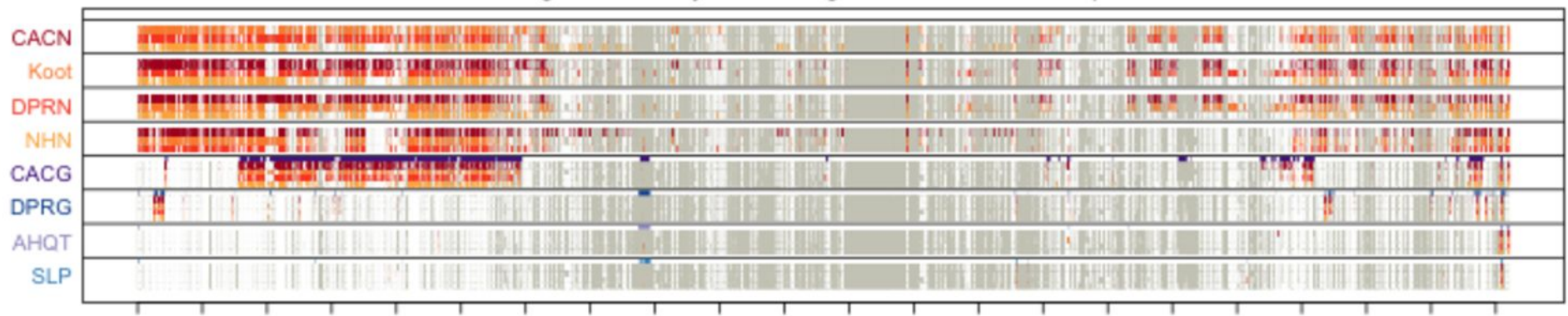
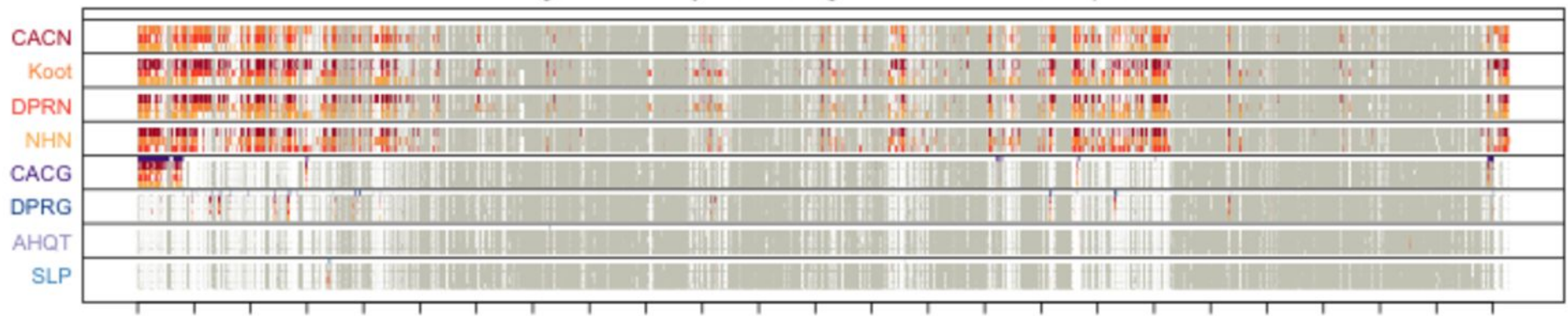

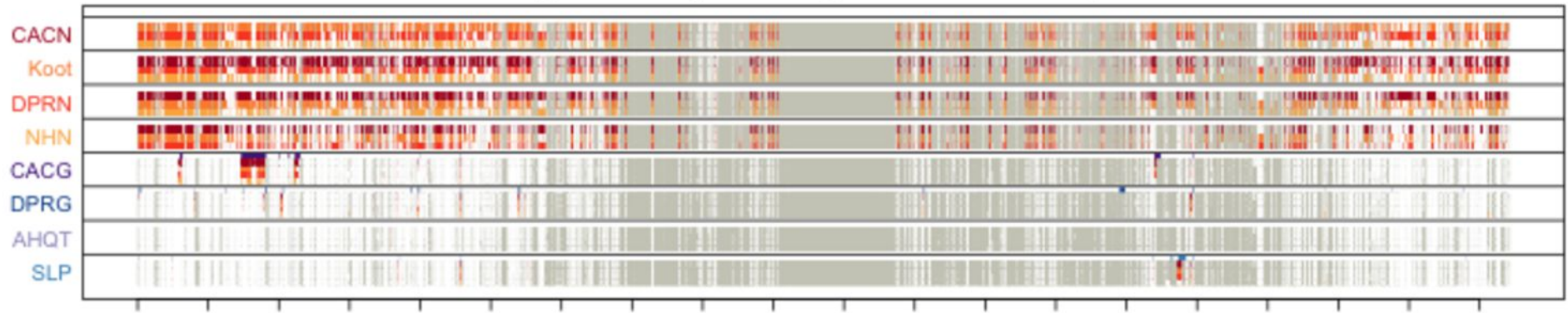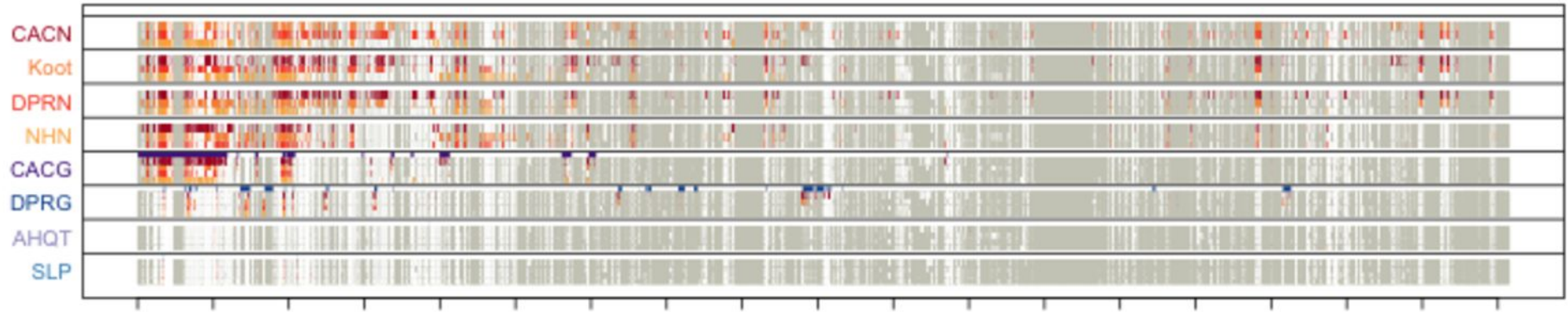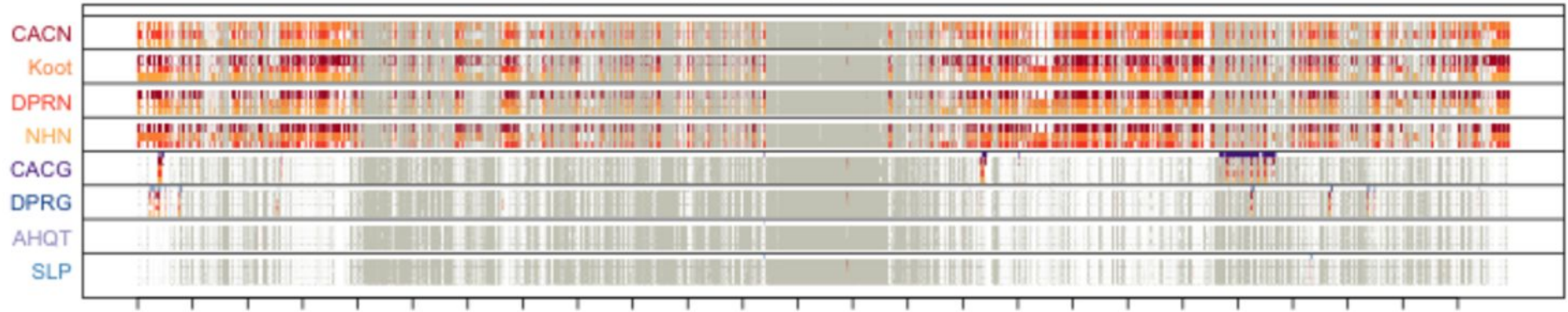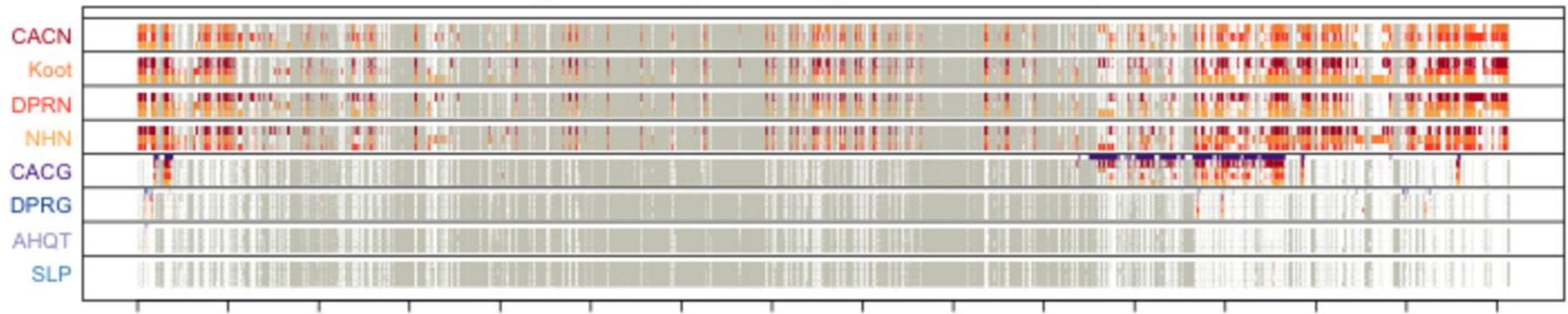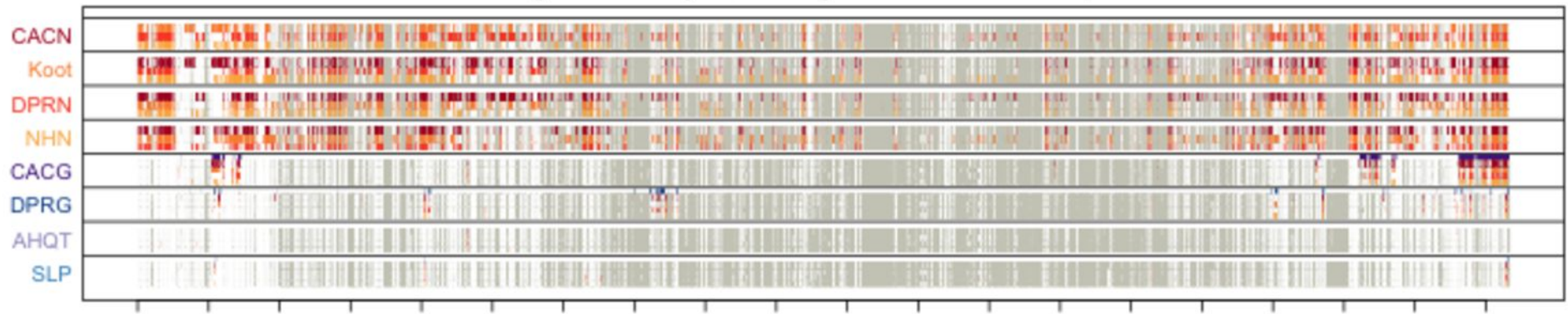

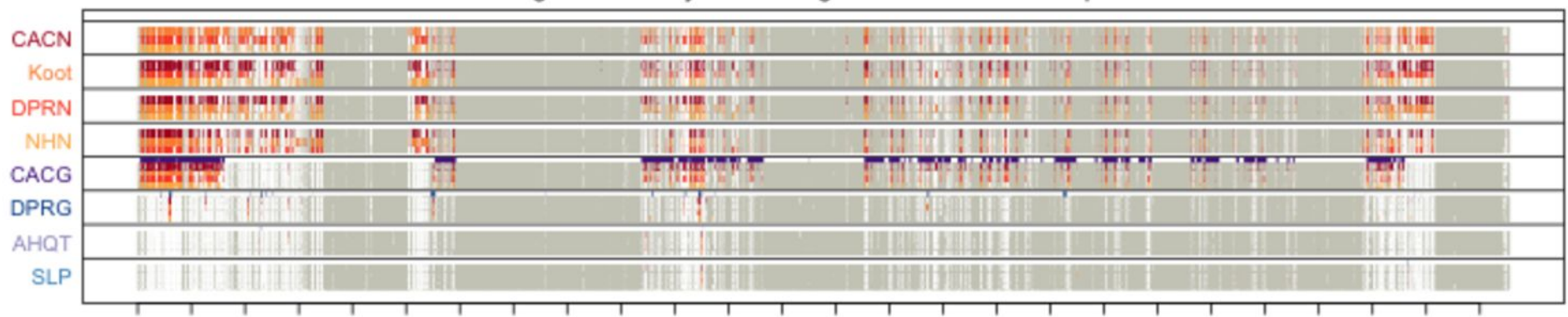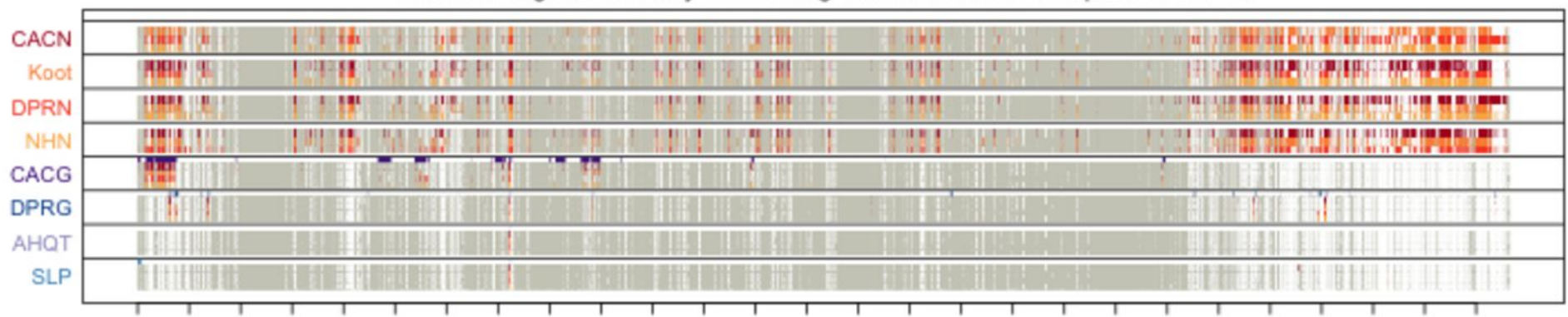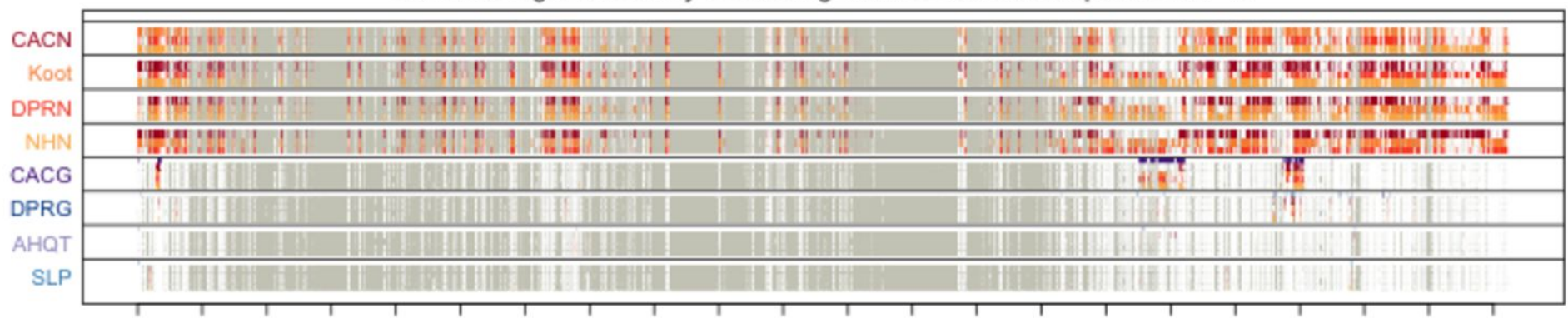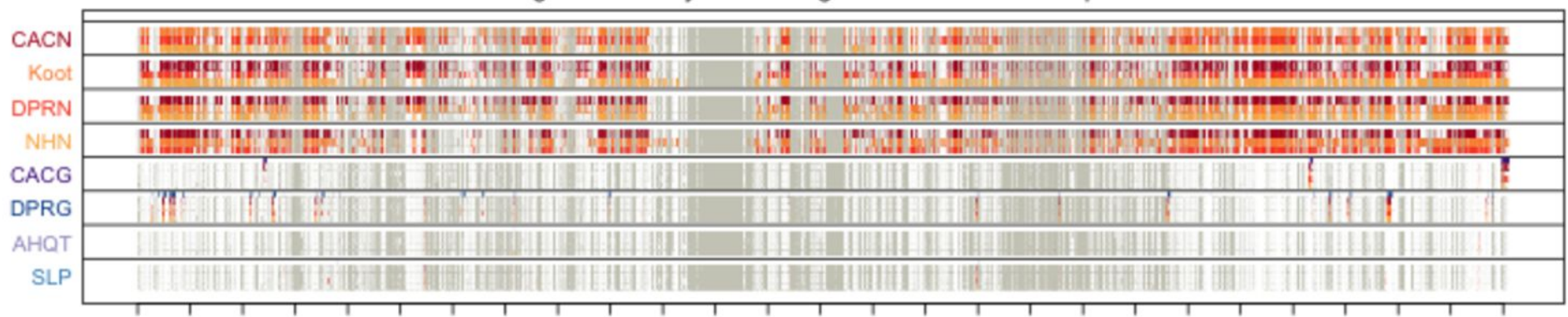

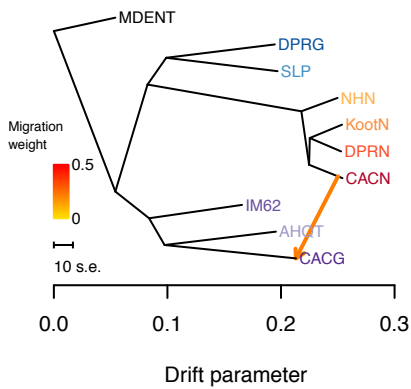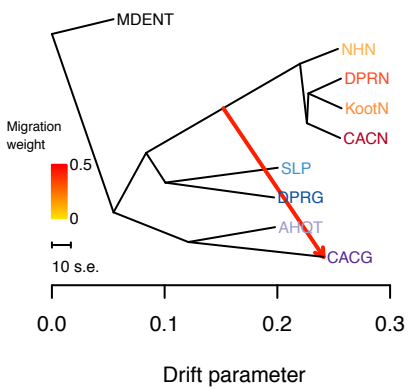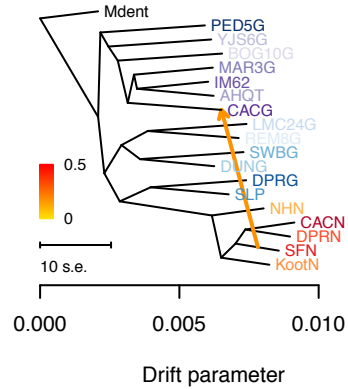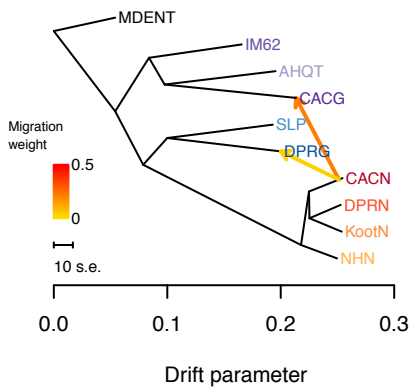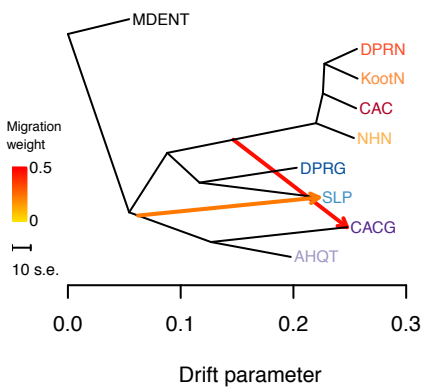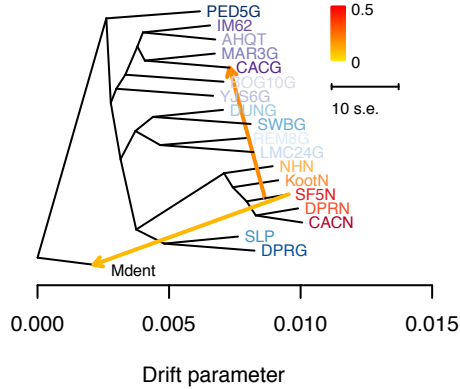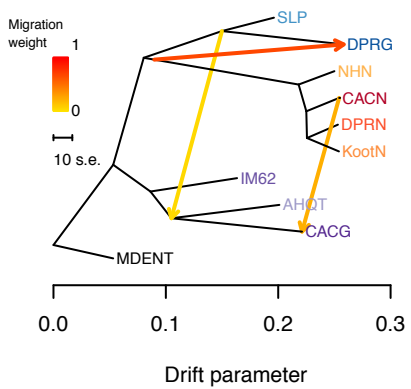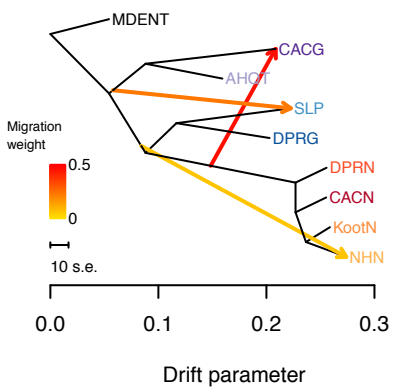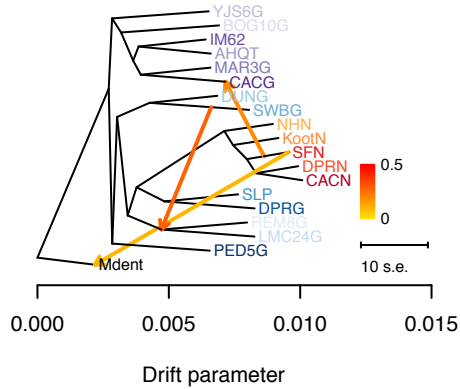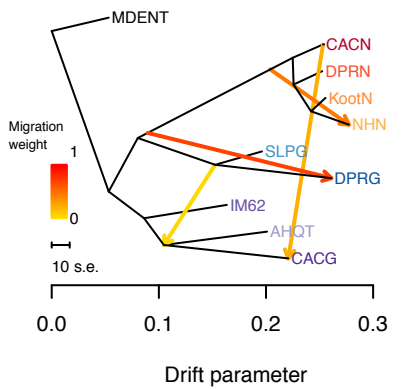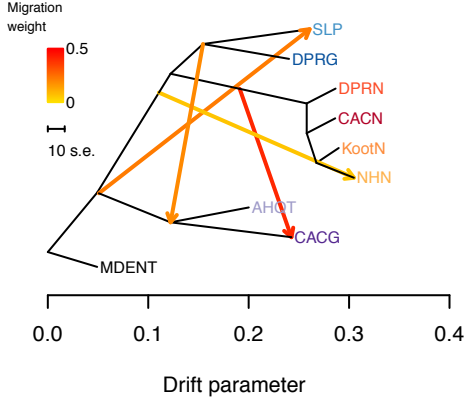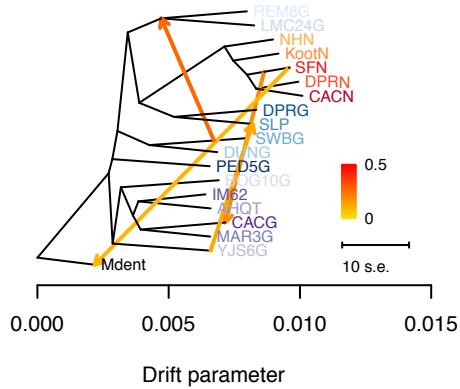

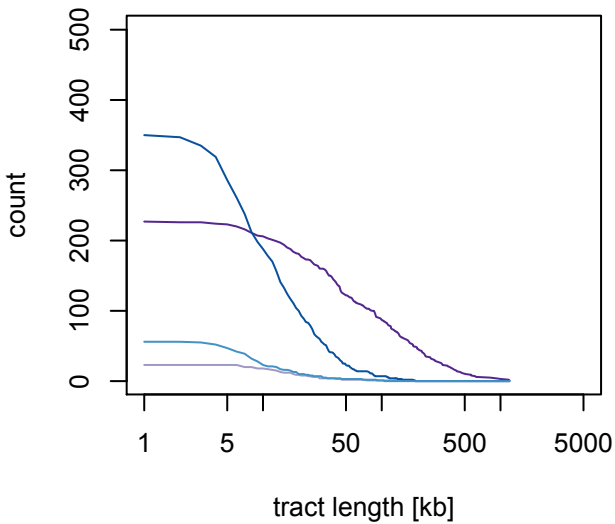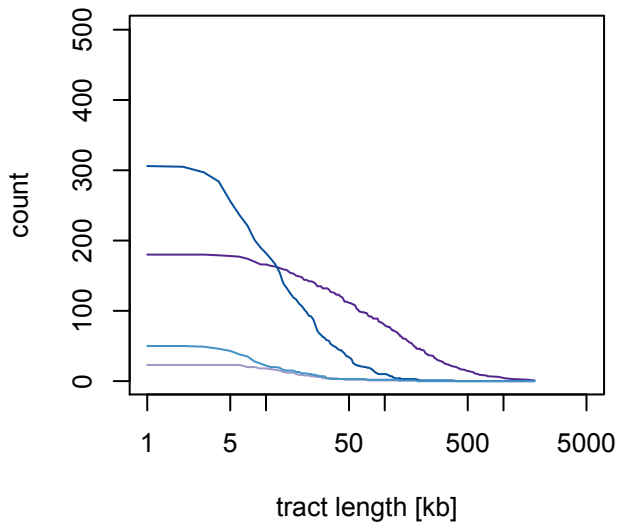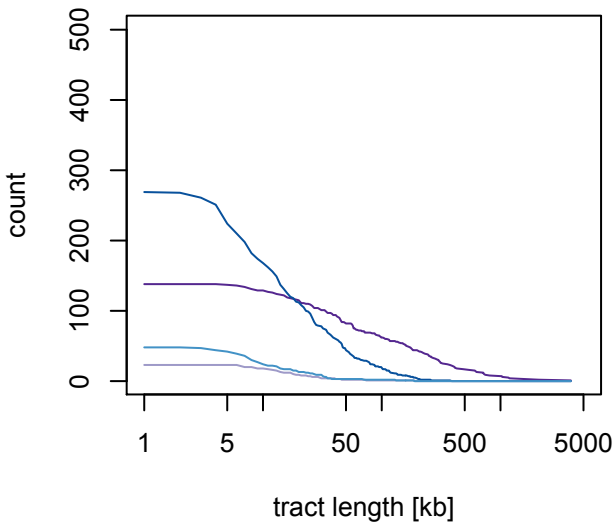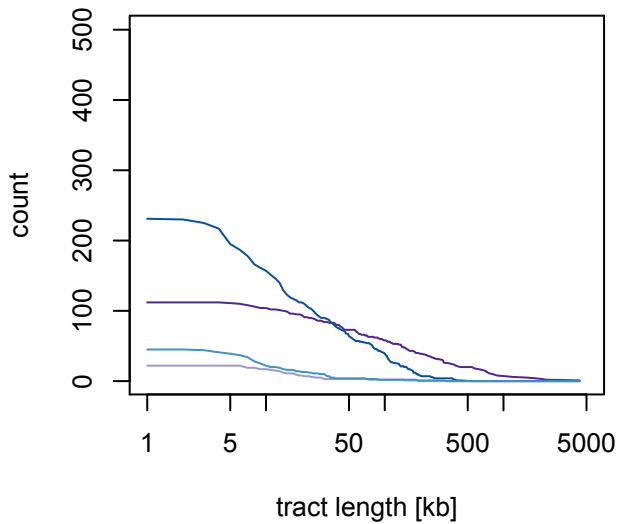